\documentclass[aps, twocolumn, prd, preprintnumbers, amsmath, amssymb, amsfonts, superscriptaddress, eqsecnum, nofootinbib]{revtex4-2}
\usepackage{graphicx} 
\usepackage{amsmath,amsfonts,amssymb,mathtools,amsthm}
\usepackage[hidelinks]{hyperref}
\usepackage{xcolor}
\usepackage[normalem]{ulem}

\newcommand{\beq}{\begin{equation}}
\newcommand{\eeq}{\end{equation}}

\newcommand{\beqs}{\begin{eqnarray}}
\newcommand{\eeqs}{\end{eqnarray}}
\newcommand{\nn}{\nonumber}

\newcommand{\GeV}{\mathrm{GeV}}
\newcommand{\MeV}{\mathrm{MeV}}

\def\bal#1\eal{\begin{align}#1\end{align}}

\begin{document}

\title{
Revisiting QCD-induced little inflation 
with chiral density wave state and its implications on pulsar timing array gravitational-wave signals
}

\author{Tae Hyun Jung}
\email{thjung0720@gmail.com}
\affiliation{Particle Theory and Cosmology Group, Center for Theoretical Physics of the Universe, Institute for Basic Science (IBS), Daejeon, 34126, Korea}

\author{Seyong Kim}
\email{skim@sejong.ac.kr}
\affiliation{Department of Physics, Sejong University, 05006 Seoul, Korea}

\author{Jong-Wan Lee}
\email{j.w.lee@ibs.re.kr}
\affiliation{Particle Theory and Cosmology Group, Center for Theoretical Physics of the Universe, Institute for Basic Science (IBS), Daejeon, 34126, Korea}

\author{Chang Sub Shin}
\email{csshin@cnu.ac.kr}
\affiliation{Department of Physics and Institute for Sciences of the Universe,
Chungnam National University, Daejeon 34134, Korea}
\affiliation{Particle Theory and Cosmology Group, Center for Theoretical Physics of the Universe,
Institute for Basic Science (IBS), Daejeon, 34126, Korea}
\affiliation{School of Physics, Korea Institute for Advanced Study, Seoul, 02455, Republic of Korea}

\author{Hee Beom Yang}
\email{qja1sk@gmail.com}
\affiliation{Department of Physics and Institute of Quantum Systems,
Chungnam National University, Daejeon 34134, Korea}


\preprint{CTPU-PTC-26-09}
\begin{abstract}
We revisit QCD-induced little inflation in which the Universe begins with a large baryon chemical potential and undergoes a strong first-order QCD phase transition, generating an observable stochastic gravitational-wave background in the nano-Hz range relevant for pulsar timing array (PTA) observations.
We point out that the conventional homogeneous transition from the quark-gluon plasma phase to the hadronic gas phase faces an unavoidable difficulty in achieving the required strength of supercooling for the observed baryon density.
This motivates us to explore whether a qualitatively different phase structure at a large baryon chemical potential can alter the relation between the baryon density and the chemical potential, and thereby modify the supercooling history of the transition.
Using the nucleon-meson model with isoscalar vector mesons, we determine the critical and spinodal structure of the chiral density wave (CDW) phase in the $(\mu_B,T)$ plane. We find that the CDW phase exhibits a nontrivial structure and can remain metastable down to a low baryon density in a certain region of the parameter space. 
Taking into account the subsequent liquid-gas transition and phase separation, however, the released latent heat is too small to realize a viable QCD-induced little inflation scenario and its associated PTA-scale gravitational-wave signal. 
Our analysis sharpens the conditions under which QCD phase transitions may act as cosmological sources of nano-Hz gravitational waves, while clarifying the possible cosmological relevance of inhomogeneous QCD phases.
\end{abstract}

\maketitle

\section{Introduction}

Strong evidence of a stochastic gravitational-wave (GW) signal was recently observed in various pulsar timing array (PTA) collaborations, NANOGrav\,\cite{NANOGrav:2023gor}, EPTA\,\cite{EPTA:2023fyk}, PPTA\,\cite{Reardon:2023gzh}, and CPTA\,\cite{Xu:2023wog}, with frequency range $1$ -- $10$ nHz.
While the signal can be explained by binary supermassive black hole inspirals, 
it is slightly favored that the signal is from other sources, such as first-order phase transitions, cosmic strings, domain walls, etc. (see e.g. Ref.\,\cite{NANOGrav:2023hvm} and references therein).
The possibility of a first-order phase transition scenario is especially interesting because the frequency range roughly coincides with the Hubble length scale of the phase transition of quantum chromodynamics (QCD).
Therefore, whether the QCD transition in the early Universe can be first order and generate an observable GW signal is an important question, and has been studied for a long time\,\cite{Witten:1984rs, Boeckel:2009ej, Boeckel:2011yj, Schettler:2010dp, McInnes:2015hga, Ahmadvand:2017xrw, He:2023ado, Han:2023znh, Shao:2024ygm, Shao:2024dxt, Schwarz:2009ii, Caprini:2010xv, Wygas:2018otj, Gao:2021nwz, Gao:2024fhm}.

In standard cosmology with a tiny baryon-to-photon ratio $\eta_B=n_B/n_\gamma\sim 10^{-9}$, the QCD transition is known to be a crossover\,\cite{Aoki:2006br}, which seems to immediately rule out the idea of explaining the PTA GW signal by the QCD phase transition.
However, as argued in Refs.~\cite{Boeckel:2009ej, Boeckel:2011yj}, one can consider a little inflationary scenario driven by the QCD phase transition, where the Universe begins with a large baryon number density after the primary inflation and undergoes a strong supercooling period before the first-order QCD phase transition terminates.\footnote{
See also Refs.~\cite{kampfer:1986,Ng:1991,Borghini_2000} for the early precursors in which the idea of QCD-driven mini/little inflation with dilution of a large initial baryon asymmetry was first discussed, and Ref.~\cite{BOECKEL2011266} for a review.}
\footnote{
Alternatively, a large lepton asymmetry can induce large chemical potentials of up and down quarks individually while keeping the net baryon density small\,\cite{Schwarz:2009ii, Wygas:2018otj}. This can lead to a first-order QCD phase transition and generate a gravitational-wave signal\,\cite{Caprini:2010xv, Gao:2021nwz}. 
We do not discuss this scenario in this work.
} 
The large baryon density could be sufficiently diluted by the latent heat released during the transition in such a way that it explains the baryon density of the current Universe. If the transition is strong and slow enough, it can also generate stochastic GWs compatible with PTA observations.

In this work, we study this \emph{QCD-induced little inflationary scenario} in more detail.
To have a large dilution factor for the observed $\eta_B$ of the current Universe, it is required that the potential barrier separating the quark-gluon plasma (QGP) phase and the hadronic phase persists down to a very low temperature and density. 
It implies a very low spinodal temperature of the QGP phase for a tiny baryon chemical potential, and we find that this is inconsistent with the known fact that the QCD transition is a crossover at a small baryon chemical potential and $T \simeq {\cal O} (100)\,\MeV$, unless the potential barrier appears below the pseudocritical temperature, which is highly unlikely. 
We discuss this point in more detail in Sec.\,\ref{sec:revisiting}.

We consider an alternative phase, the chiral density wave (CDW) phase, given that a simple picture of the QCD transition between the QGP and hadronic phases at finite temperature and density fails to accommodate the little inflationary scenario.
Unlike the conventional chiral condensate in the hadronic phase (homogeneous quark-antiquark pairing), chiral condensation in the CDW phase takes the form of particle-hole pairing and is inhomogeneous and anisotropic, forming a standing wave with a finite momentum---{\it the chiral density wave} (see Ref.\,\cite{Buballa:2014tba} for a comprehensive review).
It was suggested that such states could exist in QCD at low temperature in a certain range of large baryon density, and be favored over the color superconducting phase, arising from the formation of Cooper pairs of quarks, in the large-$N_c$ limit\,\cite{Deryagin:1992rw, Shuster:1999tn} or at finite $N_c$ with large pairing energies\,\cite{Park:1999bz}.
More specifically, a modulation between scalar and pseudoscalar condensates has been considered as a realization of the CDW in various model studies\,\cite{Nakano:2004cd, Nickel:2009wj, Frolov:2010wn, Heinz:2013hza, Carignano:2014jla, Adhikari:2017ydi,Buballa:2020xaa,Ferrer:2021mpq,TabatabaeeMehr:2023tpt,Pitsinigkos:2023xee, Papadopoulos:2024agt,Papadopoulos:2025uig}, inspired by the chiral spiral in the two-dimensional Gross-Neveu model\,\cite{Schon:2000he}. 
The CDW also provides a consistent picture of quarkyonic matter in which the confinement persists while bulk thermodynamics is dominated by a dense quark Fermi sea, with baryonic degrees of freedom governing the physics near the Fermi surface\,\cite{McLerran:2007qj, Kojo:2009ha}. 
Although there have been extensive discussions on its phenomenological implications on compact stars\,\cite{Tatsumi:2014cea, Buballa:2015awa, Carignano:2015kda, Ferrer:2021mpq, Papadopoulos:2024agt}, to the best of our knowledge, this is the first work studying the CDW phase in the context of cosmology and gravitational waves.

\begin{figure}
    \centering
    \includegraphics[width=0.48\textwidth]{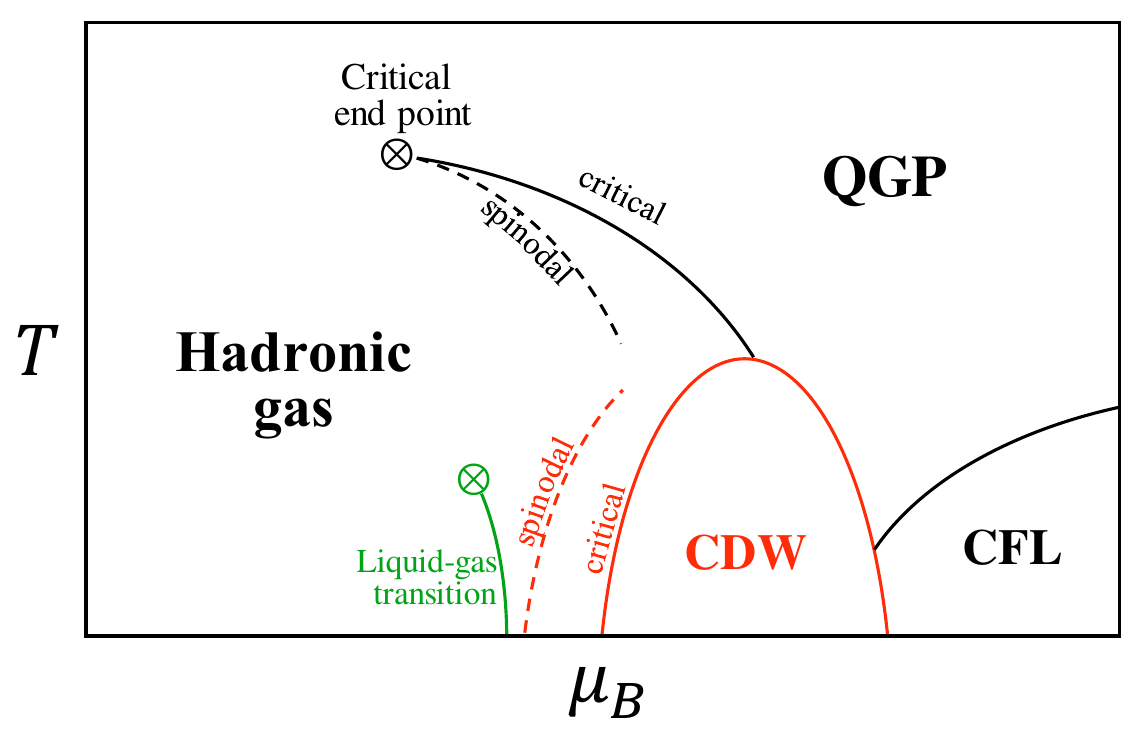}
    \caption{The schematic QCD phase diagram in the $T-\mu_{B}$ plane proposed in this work. CDW, QGP, and CFL denote the chiral density wave, quark-gluon plasma, and color-flavor locking (color superconducting) phases, respectively. }
    \label{fig:schematic}
\end{figure}

The primary goal of this work is to determine the region in the $(\mu_B,\,T)$ plane where the CDW phase is (meta)stable (see Fig.\,\ref{fig:schematic} for a schematic result) with $\mu_B$ the baryon chemical potential and $T$ the temperature.
If the metastable CDW phase can exist even for small enough $\mu_B$ and $T$ (i.e., the spinodal line of the CDW phase lies at sufficiently low baryon number density), it may provide a large dilution factor that is required in the QCD-induced little inflationary scenario. 
This is one of the key issues we investigate in this work.

The CDW phase lies in the regime that is beyond perturbative control due to the strong interaction, and first-principles lattice simulations would be required, but are not available because of the infamous sign problem; see, e.g., Ref.\,\cite{Guenther:2020jwe} for the current status of the QCD phase diagram from the lattice perspective. 
Therefore, a model study is currently the only available approach, and we employ the nucleon-meson model including (isoscalar) vector mesons~\cite{Boguta:1982wr,Floerchinger:2012xd,Drews:2013hha,Drews:2014spa}, which has been extensively studied at zero temperature for the CDW phase that may appear in the cores of a neutron star\,\cite{Pitsinigkos:2023xee,Papadopoulos:2024agt,Papadopoulos:2025uig}.
For other models, such as the quark-meson  and the Nambu-Jona-Lasinio models, see Refs.\,\cite{Nakano:2004cd, Nickel:2009wj, Frolov:2010wn, Heinz:2013hza, Carignano:2014jla, Adhikari:2017ydi,Buballa:2020xaa,Ferrer:2021mpq, TabatabaeeMehr:2023tpt}. 

As we will show, although the transition associated with the CDW phase can become strongly first-order and remain metastable down to sufficiently low density in a certain parameter region, the resulting latent heat is too small to realize a viable QCD-induced little inflationary scenario. 
This in turn requires an unrealistically low reheating temperature unable to reproduce the observed baryon yield, in tension with constraints from big bang nucleosynthesis (BBN) and cosmic microwave background (CMB). 
Consequently, we conclude that the PTA GW signal is difficult to be explained by a first-order QCD phase transition unless a more exotic phase than the CDW phase exists in the dense region of the QCD phase diagram.

This paper is organized as follows. In Sec.\,\ref{sec:revisiting}, we revisit the QCD-induced little inflationary scenario and discuss why the required phase structure is difficult to reconcile with expected properties of the conventional QCD transition at large baryon densities. 
In Sec.\,\ref{sec:model}, we introduce the nucleon-meson model with vector mesons and discuss the realization of the CDW phase in dense matter. 
In Sec.\,\ref{sec:phase_structure}, we analyze the phase structure and identify the parameter region in which the CDW phase becomes (meta)stable. 
We also study the transition from the CDW phase to the homogeneous chiral condensate and examine its implications for the little inflationary scenario. Finally, in Sec.\,\ref{sec:summary} we conclude with a summary and discussion.

\section{Revisiting QCD-induced little inflationary scenario}
\label{sec:revisiting}

The QCD-induced little inflationary scenario\,\cite{Boeckel:2009ej, Boeckel:2011yj} can be summarized as follows.
\begin{itemize}
    \item[1.] {A large $\eta_B^{(i)}$ is generated before the QCD phase transition.} 
    \item[2.] {The Universe undergoes a {\bf strong} supercooling associated with a QCD phase transition at a large baryon density.
    The supercooling dilutes the large baryon density.}
    \item[3.] {Stochastic GWs can be produced via the first-order QCD phase transition with a peak frequency $f_{\rm peak} \sim 1$\,--\,$10\,{\rm nHz}$, which may explain the GW signal observed in PTA collaborations\,\cite{NANOGrav:2023gor,EPTA:2023fyk,Reardon:2023gzh,Xu:2023wog}.}
\end{itemize}
For the first item to be fulfilled, the authors of Ref.\,\cite{Boeckel:2009ej} considered the Affleck-Dine mechanism. 
It is still questionable whether it is actually realizable, but we do not discuss this aspect because, as we argue throughout this paper, the second item does not seem to be realized by QCD in the standard model.

One may wonder if such a large $\eta_B^{(i)}$ could be consistent with BBN and CMB constraints.
Assuming that a large $\eta_B^{(i)}$ was generated in the early Universe, the QCD phase transition can become first order below $T\simeq{\cal O} (100)\,\MeV$, and a period of supercooling is expected until the bubble nucleation rate becomes comparable to the Hubble expansion rate.
During the supercooling period, the baryon density decreases as $a^{-3}$, where $a$ is the scale factor.
After the phase transition, the Universe is reheated to the reheating temperature $T_{\rm RH}$, producing large entropy and photon number densities of order $T_{\rm RH}^3$.
If $T_{\rm RH}$ is of the QCD scale, one may naively expect that a sufficiently long supercooling period can dilute the initially large baryon asymmetry.
However,  the actual condition is more restrictive once the baryon density at the end of the transition is compared directly with the entropy density after reheating.

For massless quarks with baryon chemical potential $\mu_B$ in the QGP phase, one finds
\begin{align}
n_B
=
N_c N_f
\left[
\frac{\mu_B T^2}{27}
+
\frac{\mu_B^3}{243\pi^2}
\right],
\end{align}
where we used $\mu_q=\mu_B/3$ for each quark flavor.
The entropy density receives contributions both from the relativistic thermal bath and from the $\mu_B$-dependent quark sector,
\begin{align}
s
=
\frac{2\pi^2}{45} g_{*s}^{(0)} T^3
+
\frac{N_c N_f}{27}\mu_B^2 T,
\end{align}
where $g_{*s}^{(0)}$ denotes the effective number of relativistic degrees of freedom at $\mu_B=0$.
Writing
\begin{align}
y \equiv \frac{\mu_B}{T},
\end{align}
the ratio can be expressed as
\begin{align}
\frac{n_B}{s}
=
\frac{
N_c N_f
\left(
\dfrac{y}{27}
+
\dfrac{y^3}{243\pi^2}
\right)
}{
\dfrac{2\pi^2}{45} g_{*s}^{(0)}
+
\dfrac{N_c N_f}{27} y^2
}.
\end{align}
This shows that, in the QGP phase, $\mu_B/T$ acts as the relevant variable characterizing the conserved baryon asymmetry during the cosmological evolution.
In particular,
\begin{align}
\frac{n_B}{s}\sim O(10^{-2})\,\frac{\mu_B}{T},
\end{align}
up to a numerical coefficient of order unity.

The quantity relevant for the observed baryon asymmetry is not the ratio before reheating, but the ratio after the latent heat has been converted into the radiation bath.
After reheating, the entropy density is reset to
\begin{align}
s_{\rm RH}
=
\frac{2\pi^2}{45} g_{*s}(T_{\rm RH})\, T_{\rm RH}^3,
\end{align}
where $T_{\rm RH}$ is the reheating temperature.
By baryon number conservation, the baryon density immediately after reheating can be approximated as the baryon density at the end of the supercooled stage,
\begin{align}
n_{B,{\rm RH}} = n_{B,{\rm end}},
\end{align}
if the reheating process is short.
Assuming that the system remains in the QGP phase until the end of the transition, with
\begin{align}
\mu_{B,{\rm end}} \gtrsim T_{\rm end},
\end{align}
one finds in the large-$\mu_B/T$ regime
\begin{align}
n_{B,{\rm end}}
\simeq
\frac{N_c N_f}{243\pi^2}\,\mu_{B,{\rm end}}^3.
\label{eq:n_Bend}
\end{align}
For $N_c=N_f=3$, this becomes
\begin{align}
n_{B,{\rm end}}
\simeq
\frac{1}{27\pi^2}\,\mu_{B,{\rm end}}^3
\approx
3.75\times 10^{-3}\,\mu_{B,{\rm end}}^3.
\end{align}
Therefore, the baryon-to-entropy ratio after reheating is
\begin{align}
\left.\frac{n_B}{s}\right|_{\rm RH}
=
\frac{n_{B,{\rm end}}}{s_{\rm RH}}
=
\frac{5}{6\pi^4\,g_{*s}(T_{\rm RH})}
\left(\frac{\mu_{B,{\rm end}}}{T_{\rm RH}}\right)^3.
\end{align}
This is the quantity that should be compared directly with the observed baryon asymmetry.

Taking the observed value $(n_B/s)_{\rm obs}\simeq 8.6\times 10^{-11}$, a typical reheating temperature $T_{\rm RH}\sim 100~{\rm MeV}$, and $g_{*s}(T_{\rm RH})\simeq 17.25$, the above equation implies
\begin{align}
\mu_{B,{\rm end}}
\simeq
5.6\times 10^{-3}\,T_{\rm RH}
\sim 0.5~{\rm MeV}. \label{eq:mu_Bend}
\end{align}

Thus, in the conventional homogeneous quark-gluon plasma phase, reproducing the observed baryon asymmetry after reheating requires $\mu_{B,{\rm end}}$ to be far below the typical scale of the QCD critical end point (CEP).
The crucial point is whether such a strong supercooling can be realized within standard model QCD.
As a necessary condition for this, a potential barrier between the false and true vacua in the effective potential must be maintained down to a sufficiently low $T$ and $\mu_B$.
In other words, the spinodal temperature and the corresponding chemical potential, at which the potential barrier disappears, must be extremely small compared to the QCD scale.

In Ref.\,\cite{Boeckel:2009ej}, the authors considered the dilaton-quark-meson model\,\cite{Campbell:1989gh} as an example of realizing such a case, and investigated it in more detail in Ref.\,\cite{Boeckel:2011yj}.
The key idea of the model is to introduce a dilaton field $\chi$ as an order parameter for the gluon condensate, incorporating classical scale invariance and the scale anomaly.
A classically scale-invariant potential at zero temperature and density is given by the Coleman-Weinberg potential $\sim \chi^4 \log (\chi/\chi_0)$, and it leads to its spinodal temperature zero.
At finite temperature and density, the $\chi$ field at the origin receives a quadratic correction of the form $T^2 \chi^2$ or $\mu_B^2 \chi^2$ localized around the origin, and therefore the potential barrier can persist even at a tiny $T$ or $\mu_B$.

We point out that this scenario is highly unlikely because the model fails to reproduce a crossover transition at $\mu_B/T < O(1)$.
Lattice studies disfavor a first-order phase transition in the low $\mu_B/T$ region, suggesting that the CEP, the end point of the first-order critical line in the phase diagram (see Fig.\,\ref{fig:schematic}), would be placed at $\mu_B/T \gtrsim 2$\,\cite{Giordano:2020huj} or $\mu_B>450\,\MeV$\,\cite{Borsanyi:2025dyp}.
A finite-size scaling analysis in heavy-ion colliders also puts a lower bound on $\mu_B$ of the CEP, $\mu_B \gtrsim 450\,\MeV$\,\cite{Fraga:2011hi}, while a recent work in Ref.\,\cite{Sorensen:2024mry} found evidence for a CEP near $\mu_B \simeq 625\,\MeV$ and $T\simeq 140\,\MeV$ (see also Ref.\,\cite{Lacey:2026rhc} for a related discussion).
The phase structure implied by the classical scale invariance in the dilaton-quark-meson model conflicts with the above indications.

Furthermore, a strong supercooling in the transition from QGP to the hadronic phase seems incompatible with the existence of CEP at $\mu_B/T \gtrsim 2$. This is because, in general, the spinodal and critical lines are expected to merge at the CEP, as schematically drawn in Fig.\,\ref{fig:schematic}. Consequently, the region between the spinodal and critical lines, where supercooling is allowed, cannot extend to a sufficiently small $T$ and $\mu_B$.

This is essentially the tension of the QCD-induced little inflationary scenario. 
A strong first-order transition requires the trajectory of
\begin{align}
\frac{\mu_B}{T}\gtrsim O(1),
\end{align}
and it forces $\mu_{B,{\rm end}}$ extremely small, as in Eq.~\eqref{eq:mu_Bend}, while the potential barrier at such a small $\mu_B$ is highly nontrivial.

This clarifies why it is natural to look beyond the conventional QGP-to-hadronic transition.
In the QGP phase,
the final baryon density is directly related to $\mu_{B,{\rm end}}^3$, as in Eq.~\eqref{eq:n_Bend}, so a successful dilution ultimately requires $\mu_{B,{\rm end}}$ itself to become tiny.
By contrast, in a phase with a mass gap or threshold structure, the baryon density does not need to scale simply by $\mu_B^3$, and it may become strongly suppressed even when $\mu_B$ is not parametrically small.
This motivates us to consider qualitatively different dense-QCD phases.

In particular, we focus on the regime $\mu_B/T \gg 1$, where inhomogeneous chiral condensation may become relevant. More specifically, we consider the CDW phase as a candidate initial state, where chiral condensation takes the form of fermion-hole pairing.
The system is then already in the regime of broken chiral symmetry.
Our question is whether the potential barrier that locally stabilizes such an inhomogeneous phase can persist down to the hadronic liquid-gas transition surface, across which the baryon number density drops sharply, thereby allowing the strong supercooling required in the QCD-induced little inflationary scenario.

\section{Chiral density wave in Nucleon meson model}
\label{sec:model}

To study phase transition properties of the CDW phase at $\mu_B/T \gg 1$, we consider the nucleon-meson model~\cite{Boguta:1982wr,Floerchinger:2012xd,Drews:2013hha,Drews:2014spa} (see also \cite{Glendenning:1982nc, 1985ApJ...293..470G}) 
and follow the treatment in Refs.~\cite{Pitsinigkos:2023xee,Papadopoulos:2024agt,Papadopoulos:2025uig}. 
In this section, we briefly review and summarize the model, while we consider isospin-symmetric nuclear matter with the isovector vector meson $\rho_\mu = 0$ for simplicity.

\subsection{Model setup}
The Lagrangian density of the model is decomposed into nucleonic, mesonic, and interaction terms,
\beq
    \mathcal{L} = \mathcal{L}_{\mathrm{bar}} + \mathcal{L}_{\mathrm{mes}} + \mathcal{L}_{\mathrm{int}}.
\eeq
The baryonic term is given by
\begin{equation}
    \mathcal{L}_{\mathrm{bar}} = \bar{\psi}(i\gamma^\mu \partial_\mu + \gamma^0 \mu_\psi) \psi,
\end{equation}
where $\psi$ denotes the isospin doublet nucleon (neutron $\psi_n$ and proton $\psi_p$),
\begin{equation}
    \psi=\begin{pmatrix}
\psi_p \\
\psi_n
\end{pmatrix}.
\end{equation}
For simplicity, we only consider an isospin-symmetric state, so the baryon chemical potential is given by
\beq
\mu_\psi=\begin{pmatrix}
\mu_p & 0 \\
0 & \mu_n
\end{pmatrix}
=
\begin{pmatrix}
\mu_B & 0 \\
0 & \mu_B 
\end{pmatrix}.
\eeq
Note that the nucleon mass term does not appear here, since it is dynamically generated by a phase transition order parameter, and thus will be induced from the interaction term, ${\cal L}_{\rm int}$.

The mesonic term is described by
\bal
    \mathcal{L}_{\mathrm{mes}} =
&\frac{1}{2} \partial_\mu \sigma \partial^\mu \sigma
+ \frac{1}{4}\mathrm{Tr}\left[ \partial_\mu \pi \partial^\mu \pi\right]
\label{eq:L_meson}
\\
&-\frac{1}{4}\omega_{\mu\nu}\omega^{\mu\nu}
+ \frac{1}{2} m_\omega^2 \omega_\mu \omega^\mu 
+ \frac{d}{4} (\omega_\mu \omega^\mu)^2
- \mathcal{U}(\sigma, \pi),
\nn
\eal
where $\sigma$ is the isospin singlet meson, $\pi=\pi_a\tau_a$ ($a=1,2,$ and $3$) are pions with $\tau_a$ the Pauli matrices, and 
$\omega_{\mu\nu}=\partial_\mu\omega_\nu-\partial_\nu\omega_\mu$ denotes the field strength tensor of the isoscalar vector meson field $\omega_\mu$.
We take its mass $m_\omega=782\,\MeV$, and the self-quartic interaction coupling $d\geq 0$. 
${\cal U}(\sigma, \pi)$ describes the effective potential of pseudoscalar (pion) and scalar mesons, $\pi$ and $\sigma$, and takes the form
\begin{equation}
    \mathcal{U}(\sigma,\pi)
= \sum_{n=1}^{4} \frac{a_{n}}{n!}
  \frac{\left(\sigma^{2} + \pi_{a}\pi_{a} - f_{\pi}^{2}\right)^{n}}{2^{n}}
  - \epsilon \left(\sigma - f_{\pi}\right),
  \label{eq:U_potential}
\end{equation}
where $f_\pi=93\ \mathrm{MeV}$ is the pion decay constant and $\epsilon$ is a (explicit) chiral symmetry breaking parameter.
The coefficients $a_n$ will be fixed later in Sec.\,\ref{sec:parameters}.

Finally, the interaction between baryons and mesons is given as
\begin{equation}
    \mathcal{L}_{\mathrm{int}} =
- \bar{\psi} \left[
g_\sigma (\sigma + i \gamma^5 \pi)
+  g_\omega \gamma^\mu\omega_\mu 
\right] \psi,
\end{equation}
with $g_\sigma$ and $g_\omega$, respectively, being the coupling constants of the scalar and vector mesons with nucleons.

\subsection{Order parameters and the CDW ansatz}

Chiral condensate proceeds with developing a background expectation value of the $\sigma$ field.
While the conventional hadronic phase is given by a homogeneous $\sigma$ expectation value, a modulation exists in the isospin space for the CDW phase.
A simple ansatz to describe the CDW phase has the form 
\bal
    &\sigma = \phi \cos(2\vec{q}\cdot\vec{x}),
    \nn
    \\
    &\pi_3 = \phi \sin(2\vec{q}\cdot\vec{x}),
    \label{eq:pi_3_ansatz}
    \\
    &\pi_1 = \pi_2 = 0,
    \nn
\eal
where $\vec q$ is spontaneously chosen, and we take $\phi$ as an order parameter of chiral symmetry breaking.
As one can see later, $\vec q$ is given by a microscopic scale, and therefore we average over the space and obtain the effective potential in terms of $q=|\vec q|$.

The vector meson $\omega^\mu$ also develops its background field value when $\mu_B$ is large.
Assuming spatial isotropy in the $\omega^\mu$ sector (i.e., no chiral current), we only consider the $\omega^0$ component to have its background expectation value, while $\omega^i =0$.
In the following, we denote the background field value $\omega^0$ as $\omega$ for simplicity.

Neglecting mesonic fluctuations and applying the chiral rotation to the fermionic fields,
\begin{eqnarray}
    \psi \rightarrow e^{-i \gamma^5 \tau_3 \vec{q} \cdot \vec{x}} \psi,
\end{eqnarray}
one finds the mean-field effective Lagrangian of the form
\bal
    \mathcal{L}_{\rm eff}
    &= \bar{\psi}\left( i \gamma_{\mu} \partial^{\mu} +\gamma^{0} \mu_{*} - M + \gamma^{5} \vec{q}\cdot\vec{\gamma}\,\tau_{3} \right)\psi
    \nn
    \\ 
    &\quad+ \frac{m_{\omega}^{2}}{2}\,\omega^{2}
    + \frac{d}{4}\,\omega^{4}- U - \Delta U,
    \label{eq:eff_lag}
\eal
where the effective nucleon mass is given by
\begin{equation}
    M = g_\sigma \phi,
\end{equation}
and the effective chemical potential reads
\begin{equation}
    \mu_* = \mu_B - g_\omega \omega.
\end{equation}
The mesonic vacuum potential is decomposed into an isotropic part and a $q$-dependent correction,
\bal
    &U(\phi) = \sum_{n=1}^{4} \frac{a_n}{n!} \frac{(\phi^2 - f_\pi^2)^n}{2^n} - \epsilon(\phi - f_\pi), 
    \label{eq:U}
    \\
    &\Delta U(\phi, q) = 2 \phi^2 q^2 + (1 - \delta_{q0}) \epsilon \phi.
    \label{eq:Delta_U}
\eal
The term $2\phi^2 q^2$ in $\Delta U$ originates from the scalar kinetic term in Eq.~(\ref{eq:L_meson}), whereas the term $(1 - \delta_{q0}) \epsilon \phi$ arises from the spatial average of the chiral symmetry breaking term in Eq.~\eqref{eq:U_potential} with the CDW ansatz Eq.~\eqref{eq:pi_3_ansatz}, which yields
\begin{equation}
    \overline{\sigma} =
\begin{cases}
\phi, & q = 0 \; (\mathrm{homogeneous}),\\[4pt]
0, & q \neq 0 \; (\mathrm{CDW},\ \overline{\cos}=0).
\end{cases}
\end{equation}
This treatment is valid when $q$ is much larger than the Hubble scale, which is always the case in this work.
As one can see later, the Kronecker delta term plays an important role in determining the CDW phase structure.

\subsection{Effective potential}

Now, let us compute the effective potential of the order parameters $\phi$, $q$, and $\omega$.
The baryonic contribution can be expressed as
\bal
&\Omega_{\mathrm{bar}} \label{eq:Omega_baryon}
\\
&= - 2 \sum_{e=\pm} \sum_{s=\pm} 
\int \frac{d^3\vec{k}}{(2\pi)^3} \left\{ \frac{E_k^s}{2} + T \ln\left[1 + e^{-(E_k^s - e \mu_*)/T} \right] \right\}\nn
\eal
with
\bal
E_k^{\pm}= \sqrt{\left(\sqrt{k_\ell^2+M^2}\pm q\right)^2+k_\perp^2},\\
\vec{k}_\ell=\hat{\vec{q}}\,\hat{\vec{q}}\cdot\vec{k},\quad \vec{k}_{\perp} = \vec{k} - \hat{\vec{q}}\,\hat{\vec{q}}\cdot\!\vec{k}.
\eal
The summation indices $e$ and $s$ run for fermion/antifermion and the spin along the $\vec q$ direction, respectively.
The first term of Eq.\,\eqref{eq:Omega_baryon} corresponds to the Coleman-Weinberg potential, while the second term is the free energy density in the medium, which is effectively the negative pressure of the plasma for a given order parameter $M=g_\sigma \phi$.
Thus, the baryonic contribution can be rewritten as
\begin{eqnarray}
    \Omega_\mathrm{bar}=-2(P_\mathrm{vac}+P_\mathrm{mat}).
\end{eqnarray}
The prefactor $2$ comes from the isospin degree (i.e., proton and neutron).

$P_{\rm vac}$ can be obtained by the integral,
\begin{equation}
    P_{\mathrm{vac}} = \frac{1}{2\pi^{2}}
\sum_{s=\pm} \int_{0}^{\infty} dk_{\ell}
\int_{0}^{\infty} dk_{\perp}\, k_{\perp}\, E_{k}^{s},\label{eq:P_vac}
\end{equation}
and can be further decomposed into $q$-dependent and $q$-independent contributions 
\bal
    &[-2P_{\rm vac}]_{q-{\rm indep}} \! = \! 
    \frac{m_N^4}{96\pi^2} 
    \! \left( \! 1 \!-\! 8\frac{\phi^2}{f_\pi^2} \!-\! 12\frac{\phi^4}{f_\pi^4} \ln \frac{\phi^2}{f_\pi^2} \!+\! 8\frac{\phi^6}{f_\pi^6} \!-\! \frac{\phi^8}{f_\pi^8} \right) \!,
    \\
    &[-2P_{\rm vac}]_{q-{\rm dep}} = - \frac{q^2 M^2}{2\pi^2} \ln \frac{M^2}{\ell^2} - \frac{q^4}{2\pi^2} F(M/q),
\eal
where the nucleon vacuum mass is $m_N=939\ \mathrm{MeV}$, and
\bal
    F(y) \equiv \frac{1}{3} + \Theta(1 - &y) \Bigg[ 
- \sqrt{1 - y^2} \cdot \frac{2 + 13 y^2}{6} 
\nn
\\
&+ 2 y^2 \left( 1 + \frac{y^2}{4} \right) 
\ln \left( \frac{1 + \sqrt{1 - y^2}}{y} \right) 
\Bigg].
\eal
Here, we take the renormalization scale $\ell$ in a $q$-dependent way
\begin{eqnarray}
    \ell(q) = \sqrt{m_N^2 + (2 q)^2},
\end{eqnarray}
adopting the choice taken in Ref.\,\cite{Pitsinigkos:2023xee}.

To obtain $P_{\rm mat}$, we take two different approaches for the cases with $T=0$ and $T\neq 0$.  
In the latter, we numerically integrate the second term of Eq.\,\eqref{eq:Omega_baryon}.
On the other hand, when $T=0$, the Fermi-Dirac distribution becomes a simple step function and, therefore, analytic formulas are available.
We present its expressions below.

At zero temperature and finite density, there is no antiparticle in matter, and the logarithm in Eq.\,\eqref{eq:Omega_baryon} is nonzero only for $e=+1$. 
Thus, $P_{\rm mat}$ is given by
\bal
    P_{\mathrm{mat}} \!\!
    = \! \frac{1}{2\pi^{2}} \!
\sum_{s=\pm} \! \int_{0}^{\infty} \!\! dk_{\ell} \!
\int_{0}^{\infty} \!\! dk_{\perp}\, k_{\perp}\,
\bigl(\mu_{*} - E_{k}^{s}\bigr)\,
\Theta\bigl(\mu_{*} - E_{k}^{s}\bigr).
\label{eq:P_mat}
\eal
We evaluate the double integral analytically and find
\begin{widetext}
\bal
P_\mathrm{mat}\equiv& 
\frac{\Theta(\mu_* - q - M)}{16\pi^2} \left\{ M^2 [M^2 + 4q(q - \mu_*)] \ln \left( \frac{\mu_* - q + k_-}{M} \right) + \frac{k_-}{3} \left[ 2(\mu_*^2 - q^2)(\mu_* - q) - M^2(5\mu_* - 13q) \right] \right\}\nonumber\\
&+ \frac{\Theta(\mu_* + q - M)}{16\pi^2} \left\{ M^2 [M^2 + 4q(q + \mu_*)] \ln \left( \frac{\mu_* + q + k_+}{M} \right) + \frac{k_+}{3} \left[ 2(\mu_*^2 - q^2)(\mu_* + q) - M^2(5\mu_* + 13q) \right] \right\}\nonumber\\
&+ \frac{\Theta(q - \mu_* - M)}{16\pi^2} \left\{ M^2 [M^2 + 4q(q - \mu_*)] \ln \left( \frac{q - \mu_* + k_-}{M} \right) - \frac{k_-}{3} \left[ 2(\mu_*^2 - q^2)(\mu_* - q) - M^2(5\mu_* - 13q) \right] \right\}\nonumber\\
&- \frac{\Theta(q - M)}{8\pi^2} \left[ M^2(M^2 + 4q^2) \ln \left( \frac{q + \sqrt{q^2 - M^2}}{M} \right) - \frac{\sqrt{q^2 - M^2}}{3} q (2q^2 + 13M^2) \right],
\label{eq:P_mat_zero_temp}
\eal
\end{widetext}
where $k_{\pm} \equiv \sqrt{(\mu_* \pm q)^2 - M^2}$. 
Recall that $M=g_\sigma \phi$.

In summary, the effective potential is given by
\bal
    \Omega(\phi,\omega,q,\mu_*) = &\frac{m_{\omega}^{2}}{2}\,\omega^{2}
    + \frac{d}{4}\,\omega^4
    +U+\Delta{U} \\
    &-[2P_{\rm vac}]_{q-{\rm dep}}-[2P_{\rm vac}]_{q-{\rm indep}}-2P_\mathrm{mat}.
    \nn
\eal
When $T=0$, we take $P_{\rm mat}$ from Eq.\,\eqref{eq:P_mat_zero_temp}, while we numerically integrate the second term of Eq.\,\eqref{eq:Omega_baryon} for $T\neq 0$.

\subsection{Model parameters}
\label{sec:parameters}

\subsubsection{$\epsilon$, $a_1$, and $g_\sigma$}
We first consider the vacuum state ($\mu_B=0$ and $T=0$), where we must have $\phi=f_\pi$, $q=0$ and $\omega=0$, which implies
\bal
\left. U'(\phi) \right|_{\phi=f_\pi} = a_1 f_\pi - \epsilon = 0.
\eal
In addition, after taking account of the isospin symmetry, we may reinstate the meson fluctuation by replacing $\phi^2 \to (f_\pi + \sigma)^2 + \pi^2$, and find 
\bal
a_1 = m_\pi^2 \text{~~and~~} \epsilon=f_\pi m_\pi^2.
\eal
Throughout this work, we use the physical pion mass, $m_\pi=139\,\MeV$. 
Since the nucleon mass $m_N$ is given by $g_\sigma f_\pi$, it also fixes
\bal
g_\sigma = m_N/f_\pi,
\eal
which implies $g_\sigma \simeq 10$, numerically. 

\subsubsection{$g_\omega$ and $d$}
\label{sec:g_omega}
We now fix the baryon onset chemical potential, below which the baryon number density vanishes at zero temperature,
\bal
    \mu_0 = m_N + E_B = 922.7\,\MeV,
    \label{eq:mu_0}
\eal
with the binding energy $E_B=-16.3\,\MeV$.
At saturation ($\mu_B \to \mu_0+$), the baryon density is given by $n_B \to  n_0$ with $n_0 = 0.153~\text{fm}^{-3}$. 

Although not fixed, there are empirically expected saturation values of physical quantities: the effective nucleon mass $M\to M_0 \simeq (0.7\text{ -- }0.8)\,m_N$\,\cite{Glendenning:1997wn}, and the incompressibility (or compression modulus) $K \simeq (200$\,--\,$300)\,\MeV$\,\cite{BLAIZOT1980171,PhysRevC.38.2562,Glendenning:1997wn}, where $K= \lim_{\mu\to\mu_0+}9n_B^2 \frac{d^2(\rho/n_B)}{d n_B^2}$ for the energy density $\rho$. 
These two quantities, $M_0$ and $K$, are taken as free parameters in our study.

From the gap equation along the $\omega$ field with $\mu_*=\mu_B - g_\omega \omega$,
\bal
    \frac{\partial}{\partial \omega}\Omega(\phi,\omega,q,\mu_B-g_\omega\omega) = 0,
\eal
one can find
\bal
    m_{\omega}^{2}\,\omega + d\,\omega^{3}
    - g_\omega n_0 = 0,
    \label{eq:omega_equation}
\eal
where we have used the relation 
\bal
   \frac{\partial \Omega}{\partial \mu_B}=n_B,
\eal
and we take $n_B=n_0$.
Let us denote the solution of \eqref{eq:omega_equation} as $\omega_0$, whose analytic expression will appear shortly.

\begin{table*}[t]
\centering
\caption{%
Model parameters, appearing in the effective Lagrangian of Eq.~\eqref{eq:eff_lag}, determined from given sets of selected input parameters, $M_0,~d,~K$ and $m_\pi=139\,\MeV$. 
For each set of parameters we also present the resulting values of the $\sigma$-meson mass, $m_\sigma$, and the slope parameter, $L$, at the nuclear symmetry energy of $S=32\,\MeV$.  
}
\label{tab:parameters}
\begin{tabular}{|c|c|c||c|c|c|c||c|c|}
\hline
\multicolumn{3}{|c||}{Input parameters} & \multicolumn{4}{c||}{Model parameters} &
\multicolumn{2}{c|}{Model output}\\
\hline
$M_{0}/m_{N}$ & $d$ & $K$ [MeV]  & $g_{\omega}$ & $a_{2}$ & $a_{3}$ [MeV$^{-2}$] & $a_{4}$ [MeV$^{-4}$] & $L$ [MeV] & $m_{\sigma}$ [MeV] \\
\hline\hline
0.81 & 0        & 250 & 7.87 & 52.39 & $-2.64\times10^{-2}$ & $6.19\times10^{-5}$ & 87.66 & 687.35 \\
\hline
0.81 & $10^{4}$ & 250 & 12.45 & 128.48  & $4.34\times10^{-1}$  & $7.84\times10^{-4}$ & 53.85 & 1063.26 \\
\hline
0.81 & $10^{4}$ & 300 & 12.45 & 168.26  & $5.94\times10^{-1}$  & $9.99\times10^{-4}$ & 53.85 & 1214.33 \\
\hline
0.70 & $10^{4}$ & 250 & 18.83 & 67.59   & $1.55\times10^{-1}$  & $2.00\times10^{-4}$ & 56.41 & 777.13 \\
\hline
0.89 & $10^{4}$ & 250 & 6.73  & 376.87  & 2.24 & $5.81\times10^{-3}$ & 55.77 & 1810.76 \\
\hline
0.93 & $10^{4}$ & 250 & 2.81  & 2451.31 & 24.69 & $8.77\times10^{-2}$ & 64.92 & 4606.60 \\
\hline
\end{tabular}
\end{table*}

Since we consider the Fermi-Dirac distribution at $T=0$ with $q=0$, $\mu_{*0}$, the saturation value of $\mu_*$, is solely determined by $M_0$ as
\bal
\mu_{*0}\equiv \mu_0-g_\omega \omega_0 =\sqrt{k_{F0}^2 +M_0^2} 
\eal
where $k_{F0} = (3\pi^2 n_0/2)^{1/3} \simeq 259\,\MeV$ is the Fermi momentum at saturation.
Inserting $\omega_0 = (\mu_{*0}-\mu_0)/g_\omega$ into \eqref{eq:omega_equation}, we obtain an equation for $g_\omega$, whose solution is given by
\bal
g_{\omega}^{2}
    = \frac{m_{\omega}^{2}}{2n_{0}}\,
      (\mu_{0}-\mu_{*0})
      \left[
      1 + \sqrt{1 + \frac{4d n_{0}(\mu_{0}-\mu_{*0})}{m_{\omega}^{4}}}
      \right].
\label{eq:g_omega}
\eal
This fixes $g_\omega$ for each $M_0$ and $d$.
To make $g_\omega$ real, we need $\mu_{*0}<\mu_0$, implying an upper bound of $M_0$
\bal
M_0 < \sqrt{\mu_0^2 - k_{F0}^2} \simeq 0.943\,m_N,
\label{eq:M0_theoretical_upper}
\eal
which restricts our choice of $M_0$. 

The solution of \eqref{eq:omega_equation} in $\omega$ is given by
\bal
    \omega_{0} = \frac{g_{\omega} n_{0}}{m_{\omega}^{2}}\,f(x_{0}),\label{eq:omega_0}
\eal
where
\bal
    &f(x) \equiv \frac{3}{2x}\,
    \frac{1 - \bigl(\sqrt{1+x^{2}} - x\bigr)^{2/3}}
         {\bigl(\sqrt{1+x^{2}} - x\bigr)^{1/3}},
    \\
    &x_{0} \equiv \frac{3\sqrt{3d}\,g_{\omega} n_{0}}{2m_{\omega}^{3}}.
\eal
As $f(x) \simeq 1 -\frac{4}{27}x^2 +O(x^4)$ for a small $x$, we have $\omega_0 \simeq \frac{g_\omega n_0}{m_\omega^2}(1 - \frac{g_\omega^2 n_0^2}{m_\omega^6} d +O(d^2))$. 
Given that the coefficient of $d$ is numerically $O(10^{-4})$, we may restrict $d$ less than $10^4$ such that the condensate $\omega$ should not be disturbed by $d$ too much\,\cite{Papadopoulos:2024agt}.

\subsubsection{$a_2$, $a_3$, and $a_4$}

To fix the coefficients that appear in the $\sigma$ potential in Eq.~\eqref{eq:U_potential}, we take the following three conditions.
First of all, we use the gap equation $\frac{\partial \Omega}{\partial \phi}=0$,
\bal
0 = U'(\phi_0)
+ \frac{g_{\sigma}M_{0}}{\pi^{2}}
\!\left(
k_{F}\mu_{*0} - M_{0}^{2}\ln\!\frac{k_{F}+\mu_{*0}}{M_{0}}
\right)
\eal
with $\phi_0=M_0/g_\sigma$.
Second, we require the free energy of saturated nuclear matter to coincide with that of the vacuum, 
\bal
\label{eq:second_condition}
0 =& \frac{m_{\omega}^{2}}{2}\omega_{0}^{2}
+ \frac{d}{4}\omega_{0}^{4}
- U(\phi_0)\\
&+ \frac{1}{4\pi^{2}}
\!\left[
\!\left(\frac{2}{3}k_{F}^{3} - M_{0}^{2}k_{F}\right)\mu_{*0}
+ M_{0}^{4}\ln\!\frac{k_{F}+\mu_{*0}}{M_{0}}
\!\right].
\nn
\eal
Finally, we use the definition of $K$, and obtain
\bal
0 = &U''(\phi_0)+ \frac{g_{\sigma}^{2}}{\pi^{2}}\left(\frac{k_{F}^{3} + 3k_{F}M_{0}^{2}}{\mu_{*0}}- 3M_{0}^{2}\ln\!\frac{\mu_{*0}+k_{F}}{M_{0}}\right)- \frac{3k_{F}^{2}}{\mu_{*0}}\nn\\
    &+ \frac{6g_{\sigma}^{2}k_{F}^{3}}{\pi^{2}}\!\left(\frac{M_{0}}{\mu_{*0}}\right)^{2}\!\Big/\!\left[K - \frac{6k_{F}^{3}}{\pi^{2}}\frac{g_{\omega}^{2}}{m_{\omega}^{2}}(f(x_{0})+x_{0}f'(x_{0}))\right].
\eal
Note that all these equations are linear in $a_2$, $a_3$, and $a_4$ because $U(\phi)$ or $U''(\phi)$ appear linearly.
Hence, we solve them analytically and fix the coefficients, although we do not present their expression due to its complexity.

In summary, we have fixed the model parameters, $\epsilon$, $a_1$, and $g_\sigma$, using the vacuum values of $m_\pi$, $f_\pi$, and $m_N$. The other parameters, $g_\omega$, $a_2$, $a_3$, and $a_4,$ are also determined from the gap equations with respect to $\phi$ and $\omega$, as well as certain conditions at the nuclear saturation density. The latter involves three physical quantities, $M_0$, $d$, and $K$, which are not well constrained by experiments. We therefore take these as free input parameters, appropriately chosen to realize the CDW phase without spoiling the physics of dense nuclear matter much. We are particularly generous with the upper bound of $M_0$, because it turns out that a wider parameter space for the CDW phase opens up as $M_0$ increases. 

In Table\,\ref{tab:parameters}, we present the typical values of the input parameters considered in this work and the deduced values of the model parameters, $g_\omega$, $a_2$, $a_3$, and $a_4$. The values in the third and fourth rows match some benchmark parameters used in Ref.\,\cite{Pitsinigkos:2023xee}. 
In the last two columns, we also report two physical quantities that can be determined in our model setup, which might be used as an indication for the limitation of the nucleon-meson model with the mean-field approximation. Following Ref.~\cite{Pitsinigkos:2023xee}, we first calculate the slope parameter $L$ at the fixed value of the nuclear symmetry energy $S=32\,\MeV$, and find that the resulting values are within a typical range of the experimental bounds, $L \simeq 40$\,--\,$140\,\MeV$ for $S \simeq (30$\,--\,$34)\,\MeV$. The other quantity we show in the table is the mass of the $\sigma$ meson, which is nothing but the curvature of the potential $U$ at the vacuum expectation value of $\langle \sigma \rangle =f_\pi$,
\bal
m_{\sigma}^{2} = U''(f_{\pi}) = m_{\pi}^{2} + a_{2} f_{\pi}^{2} .
\eal
The resulting values of $m_\sigma$ vary widely with the model parameters, from a few hundreds to a few thousands in MeV, which is in the right range compared to the experimental results with an order-of-magnitude estimation. 
Note that $m_\sigma$ is poorly determined by experiments, even its identification is elusive: in Particle Data Group \cite{ParticleDataGroup:2024cfk}, the pole mass ranges in $(400$\,--\,$550)\,\MeV - i(200$\,\--\,$350)\,\MeV$, while the Breit-Wigner resonance mass ranges in $400$\,--\,$800\,\MeV$ with the decay width $100$\,--\,$800\,\MeV$. 

In the following, we are not going to pay much attention to the experimental constraints on $M_0$ and $m_\sigma$.
The reason is that the primary purpose of this work is testing whether the QCD-induced little inflationary scenario can be realized by the CDW phase transition, but it is not whether the CDW phase is possible or not.
As can be seen in the next section, the CDW phase in this model appears in a large $M_0$ and $d$ region, which are already in tension with the experimental constraints on $M_0$ or $m_\sigma$.
Moreover, we would like to emphasize that any model description cannot be fully realistic.
For instance, the vector meson potential does not have to be in the polynomial form, as $\omega$ is a composite state (i.e., we could include higher-dimensional operators, or use an arbitrary special function in terms of $\omega^\mu\omega_\mu$).
Since the potential shape determines the existence of the CDW phase as well as $m_\sigma$ (see, e.g., Eq.\,\eqref{eq:second_condition}), one could design the potential to be consistent with favorable values.
This would be meaningless for the purpose of our study.
Instead of putting effort into it, we simply release the constraints on $M_0$ and $m_\sigma$, and focus on the CDW phase transition properties and their implications in the cosmological context.
Our conclusion at the end of the day is that the CDW phase cannot realize the QCD-induced little inflationary scenario even without these constraints.
This justifies our relaxation.

\section{Phase structure and transition properties}
\label{sec:phase_structure}

\subsection{CDW phase}
\label{sec:CDW_phase}

\begin{figure*}[t]
    \centering
    \includegraphics[width=0.32\textwidth]{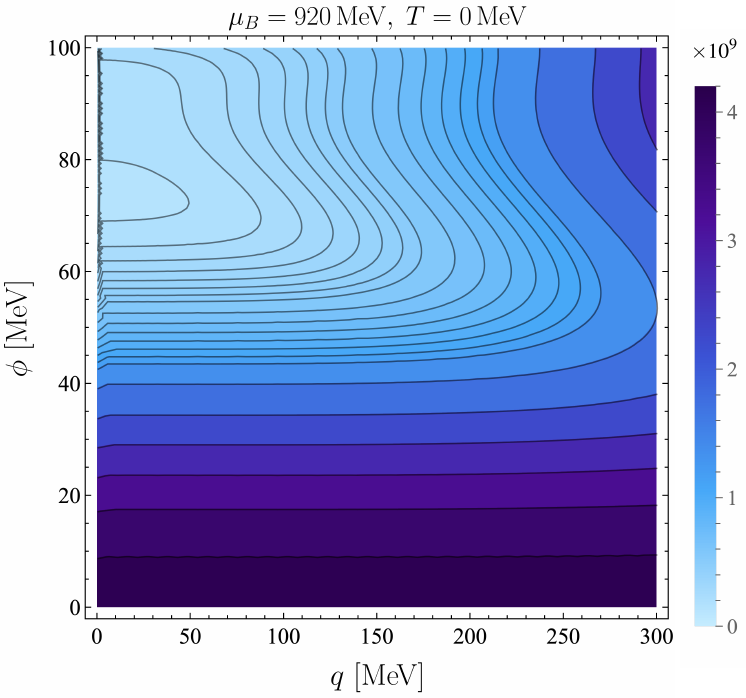}
    \includegraphics[width=0.32\textwidth]{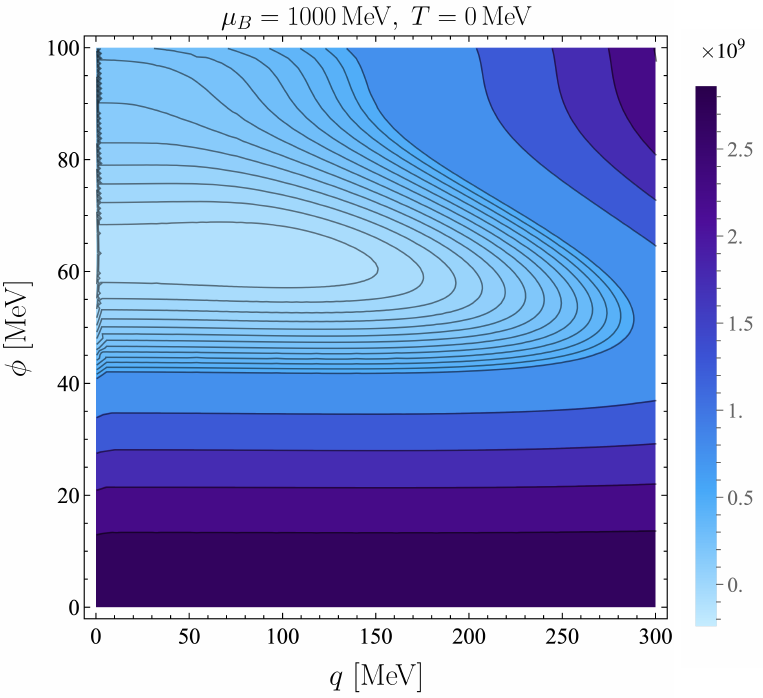}
    \includegraphics[width=0.32\textwidth]{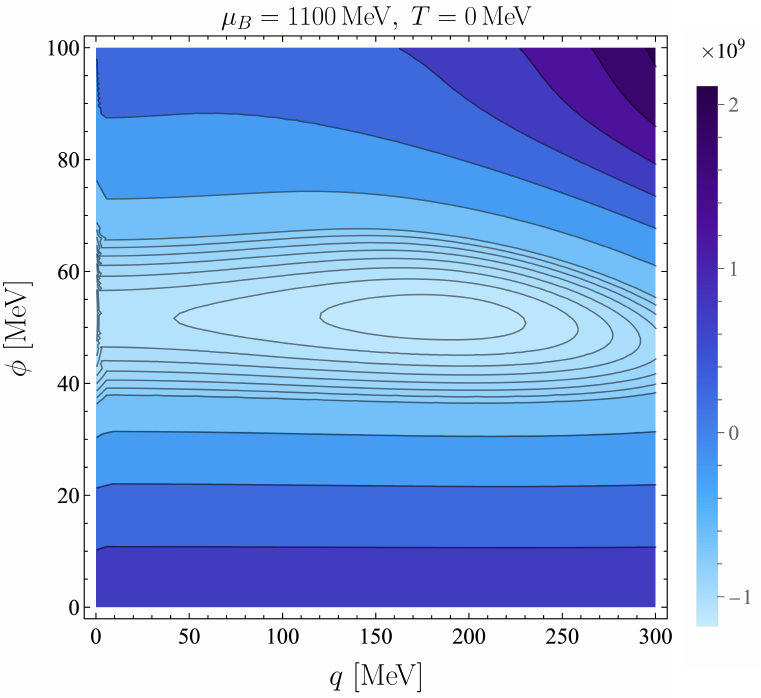}
    \caption{Contour plots of the thermodynamic potential $\Omega(T,\mu_B,\phi,q)$ in the $(\phi,q)$ plane for various values of $\mu_B$ and $T=0$, with $M_0=0.81m_N$, $K=250$ MeV, and $d=10^4$. The middle and right panels show the emergence of a local minimum at a nonzero $q$. These plots do not depict the potential at $q=0$.}
    \label{fig:potential_zero_T}
\end{figure*}

\begin{figure}[t]
    \centering
    \includegraphics[width=0.48\textwidth]{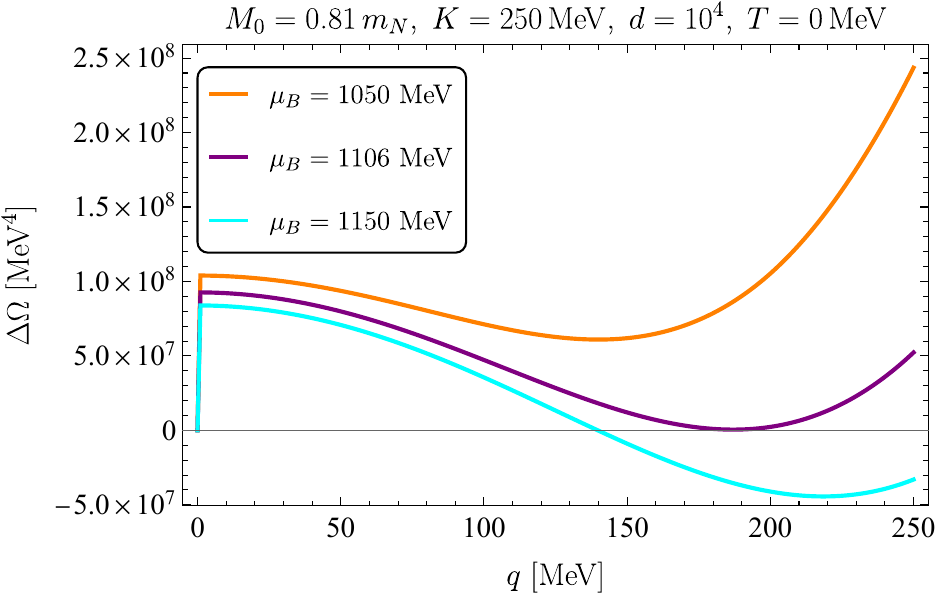}
    \caption{The reduced effective potential $\Delta \Omega$, defined in Eq.~\ref{eq:reduced_poten}, as a function of the wave number $q$ for various values of the chemical potential $\mu_B$ at $T = 0$. The other input parameters are fixed to $M_0=0.81m_N$, $K=250\,\MeV$, $d=10^4$. }
    \label{fig:potential_in_q_zero_T}
\end{figure}

\begin{figure}[t]
    \centering
    \includegraphics[width=0.48\textwidth]{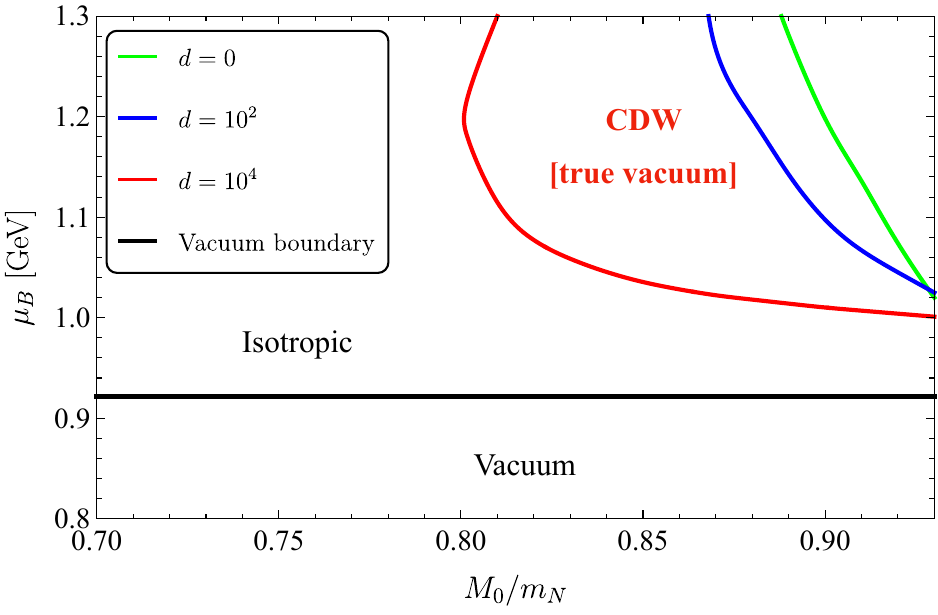}
    \caption{Zero-temperature phase structure as a function of the model parameter $M_0$. We fix $K=250$ MeV and use three representative values of $d$ as shown in the legend. The black solid line denotes the baryon onset chemical potential $\mu_0=922.7\,\MeV$ below which $n_B=0$.}
    \label{fig:CDW_phase_in_M0-muB}
\end{figure}

The wave number $q$ serves as an order parameter for the CDW phase---the CDW phase exists when a local minimum of the free energy density $\Omega$ is located at a nonzero $q$.
We investigate it by solving the gap equation with respect to the wave number $q$,
\begin{align}
    \frac{\partial \Omega}{\partial q}=0,
\end{align}
while simultaneously solving the other gap equations $\frac{\partial \Omega}{\partial \phi}=\frac{\partial \Omega}{\partial \omega}=0$.

On the other hand, the homogeneous chiral condensate ($q=0$) always exists due to the Kronecker delta term in Eq.~\eqref{eq:Delta_U}.
Thus, we compare local minima at $q=0$ and $q \neq 0$ (if it exists) to see the stability of the CDW phase.

Let us first investigate the zero-temperature case, which was also studied in Refs.~\cite{Pitsinigkos:2023xee, Papadopoulos:2024agt,Papadopoulos:2025uig} in various contexts.
Later, we will consider finite temperature, and obtain the CDW phase diagram in the $(\mu_B,\,T)$ plane for the first time in this model.

In Fig.\,\ref{fig:potential_zero_T}, we show the potential behavior in the $q$ and $\phi$ plane at zero temperature for $M_0=0.81m_N$, $K=250\,\MeV$, and $d=10^4$.
Here, we take $\omega$ to minimize $\Omega$ at each value of $q$ and $\phi$, while we do not depict the potential values at $q=0$ due to the visualization issue of the Kronecker delta.
As one can see in the middle and right panels ($\mu_B \gtrsim 1\,\GeV$), a local minimum at a nonzero $q$ emerges.
Note that in these local minima, $\phi$ values are also slightly lowered.

To compare these minima with the potential value at $q=0$, we define the reduced effective potential as
\bal 
\Delta \Omega (q) = \Omega (\phi_{\rm min}(q), \omega_{\rm min}(q), q) -\Omega(\phi_{\rm min}(0), \omega_{\rm min}(0),0),
\label{eq:reduced_poten}
\eal
where $\phi_{\rm min}(q)$ and $\omega_{\rm min}(q)$ are field values of $\phi$ and $\omega$ minimizing the potential for each $q$. We present the results in Fig.\,\ref{fig:potential_in_q_zero_T}, at three representative values of $\mu_B$ denoted by different colors. We find that the critical chemical potential is around $\mu_B^c = 1106\,\MeV$ corresponding to the purple curve.
At $\mu_B > \mu_B^c$, the CDW phase is energetically more favored than the hadronic phase with the homogeneous chiral condensate. 
Even for $\mu_B<\mu_B^c$, the CDW phase can exist but becomes metastable (see the orange curve for $\mu_B=1050\,\MeV$). 
When the (meta)stable CDW phase exists, the potential barrier between the two minima is provided by the Kronecker delta term in Eq.~\eqref{eq:Delta_U} and the decreasing behavior of the potential with $q>0$.

Following the discussion above about the identification of a stable CDW phase, we scan the parameter space by varying $M_0$ while keeping the other input parameters fixed. The results at zero temperature are shown in Fig.~\ref{fig:CDW_phase_in_M0-muB}. We have considered three representative values of $d=0$ (green), $10^2$ (blue) and $10^4$ (red) to illustrate its dependence of the CDW phase, while we have fixed $K=250\,\MeV$. In the figure, we also present the baryon onset chemical potential $\mu_0=922.7\,\MeV$ by a solid black line below which $n_B=0$. 
We find that the CDW phase exists over a wide range of parameter space and the corresponding region expands as the $d$ value increases. The latter indicates that the isoscalar vector meson plays a crucial role in stabilizing the CDW vacuum, and thus should be carefully taken account of for dedicated studies of the CDW phase, which is beyond the scope of this work. Our results agree well with Ref.\,\cite{Pitsinigkos:2023xee}.

\begin{figure}[t]
    \centering\includegraphics[width=0.48\textwidth]{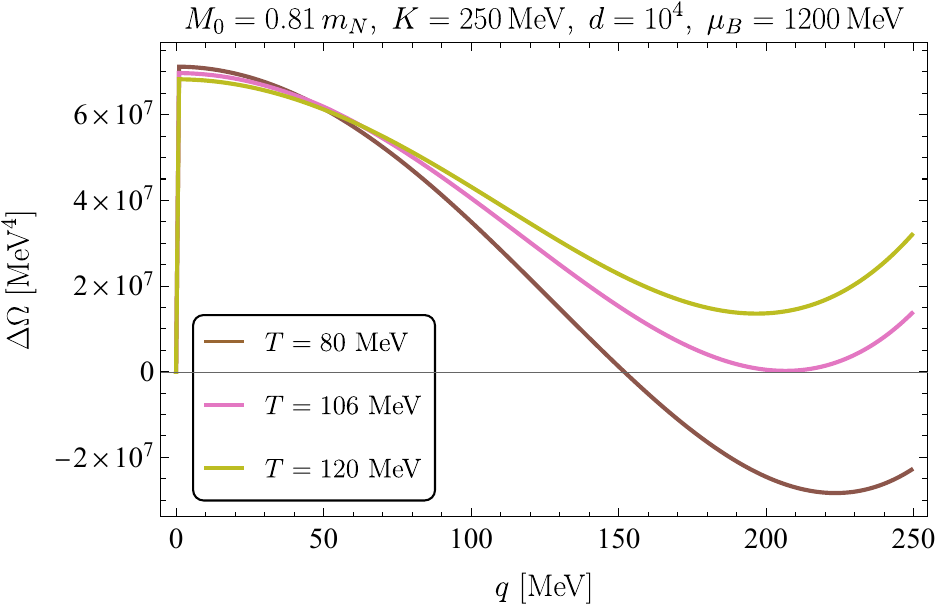}
    \caption{Free-energy difference of the CDW phase with respect to the thermodynamically stable isotropic phase, $\Delta \Omega$ in Eq.~(\ref{eq:reduced_poten}), as a function of temperature $T$ at fixed $\mu_B = 1200$ MeV. 
    The input parameters are fixed to $M_0=0.81m_N$, $K=250$ MeV, and $d=10^4$. }
    \label{fig:potential_in_q_nonzero_T}
\end{figure}

\begin{figure}[t]
    \centering
    \includegraphics[width=0.48\textwidth]{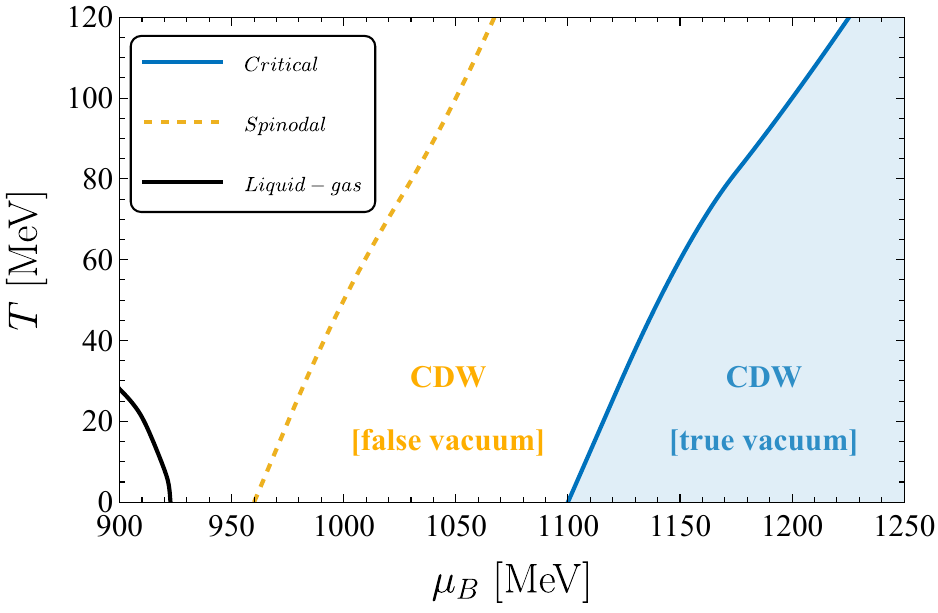}
    \caption{Phase diagram of dense nuclear matter in the $\mu_B$-$T$ plane. The blue solid and orange dashed lines denote the critical and spinodal curves below which the CDW phase is stable and metastable, respectively. We also present the liquid-gas transition by the black solid line. The input parameters are fixed to $M_0=0.81m_N$, $K=250$ MeV, and $d=10^4$.}
    \label{fig:CDW_phase_diagram}
\end{figure}

Now, let us discuss the behavior when the temperature is nonzero.
In Fig.\,\ref{fig:potential_in_q_nonzero_T}, we show the temperature dependence of the reduced effective potential $\Delta \Omega$ at a fixed $\mu_B=1200\,\MeV$.
As the temperature increases, it shows that the CDW phase becomes less stable, i.e. the local minimum of the potential at $q\neq 0$ is lifted up. The critical temperature can be determined from the condition,  $\Delta\Omega(q_{\rm min})|_{T=T_c}=\Delta\Omega(0)=0$ with $q_{\rm min}$ satisfying  $\Delta\Omega'(q_{\rm min})=0$, which yields $T_c\simeq 106\,\MeV$ in this example. 
By extending this analysis to the region of $900\,\MeV\leq\mu_B\leq 1250\,\MeV$ and $0\,\MeV\leq T\leq 120\,\MeV$, while keeping the other input parameters fixed by $M_0=0.81\, m_N$, $K=250\,\MeV$, and $d=10^4$, we find the phase diagram in the $\mu_B$-$T$ plane, as shown in Fig.\,\ref{fig:CDW_phase_diagram}. There are two reasons why we restrict ourselves to the aforementioned region of the parameter space. First of all, this is the region in which the validity of the nucleon-meson model is guaranteed. More importantly, we recall that our interests are in the transition between the CDW and hadronic phases with $\mu_B/T\ll 1$, as discussed in Sec.\,\ref{sec:revisiting}. 

The critical line depicted in the figure is one of the most distinct features of our model, which is significantly different from the typical results in the literature. 
For instance, the quark-meson (QM) model without a vector meson contribution has an opposite tendency; the CDW phase is more stable when the temperature increases in a similar range of $\mu_B$, e.g. see Refs.\,\cite{Nickel:2009wj,Carignano:2014jla,Adhikari:2017ydi,Buballa:2020xaa}. Even if we do not show the results, we should note that the NM model with $d_\omega =0$ exhibits a qualitatively similar behavior to the QM model. We can therefore conclude that the $\omega$ meson with a large $d$ is responsible for this distinguishable feature.

\subsection{Spinodal line of the CDW phase}

We now consider a cosmological evolution starting from a CDW phase\footnote{One may consider an even larger initial chemical potential and start from a color superconducting phase, and transition into the CDW phase. But, this is beyond the scope of this work.}.
As the Universe expands, $\mu_B$ and $T$ decrease. 
When they fall below the critical line of the first-order phase transition, the hadronic phase at $q=0$ becomes more stable, enabling thermal/quantum tunneling via bubble nucleation.
Initially, the bubble nucleation rate is low, so the phase transition does not proceed because the space-time expansion in the metastable CDW phase is more rapid than the bubble's nucleation and its growth.
Thus, supercooling will last until the bubble nucleation rate becomes comparable to the Hubble rate.

\begin{figure*}[t!]
    \centering
    \includegraphics[height=5.0cm]{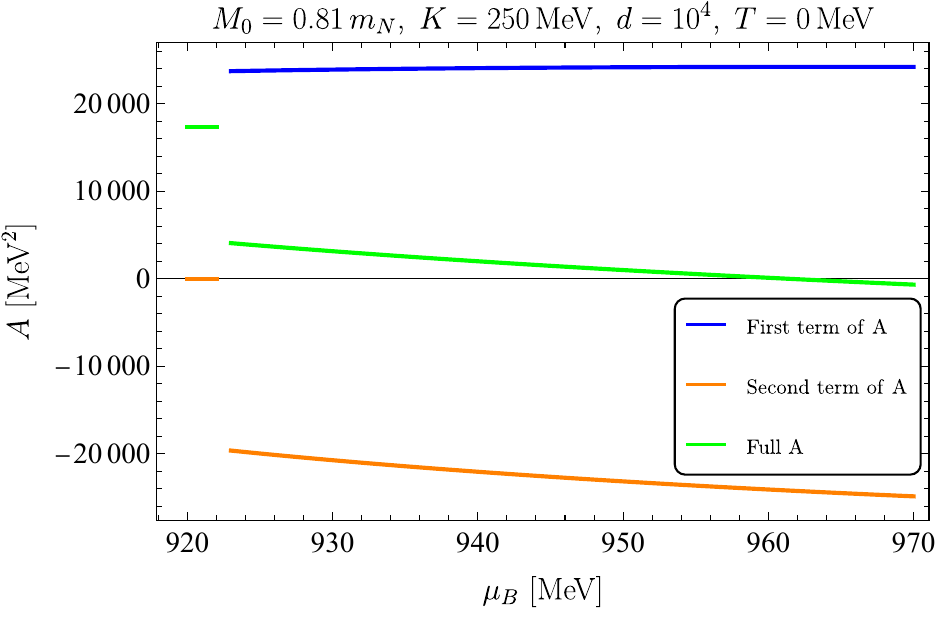}\includegraphics[height=5.6cm]{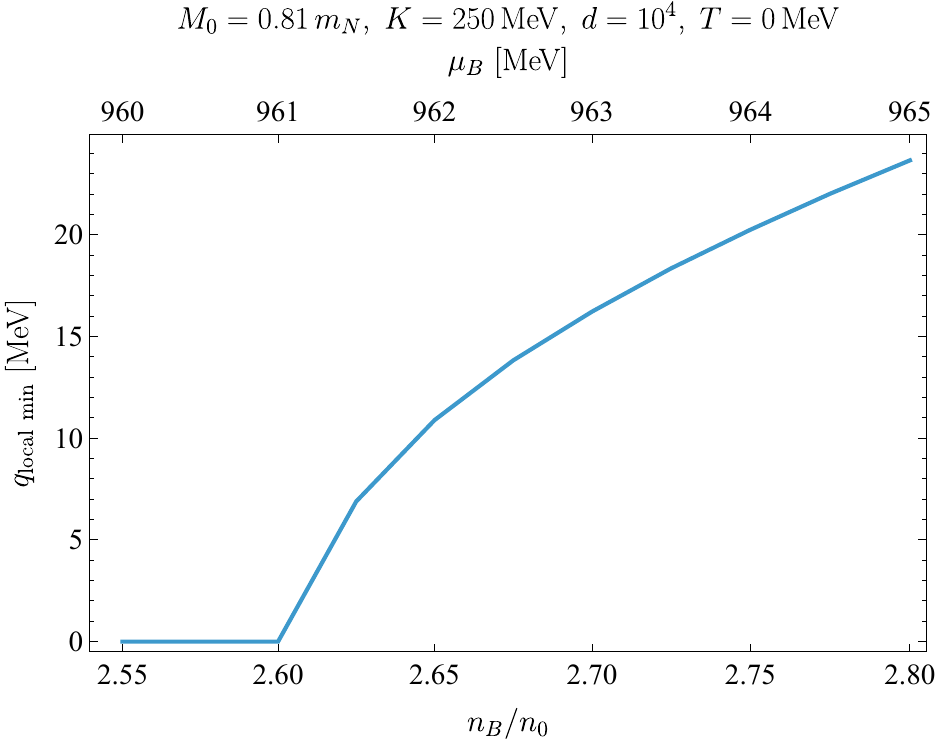}
    \includegraphics[height=5.0cm]{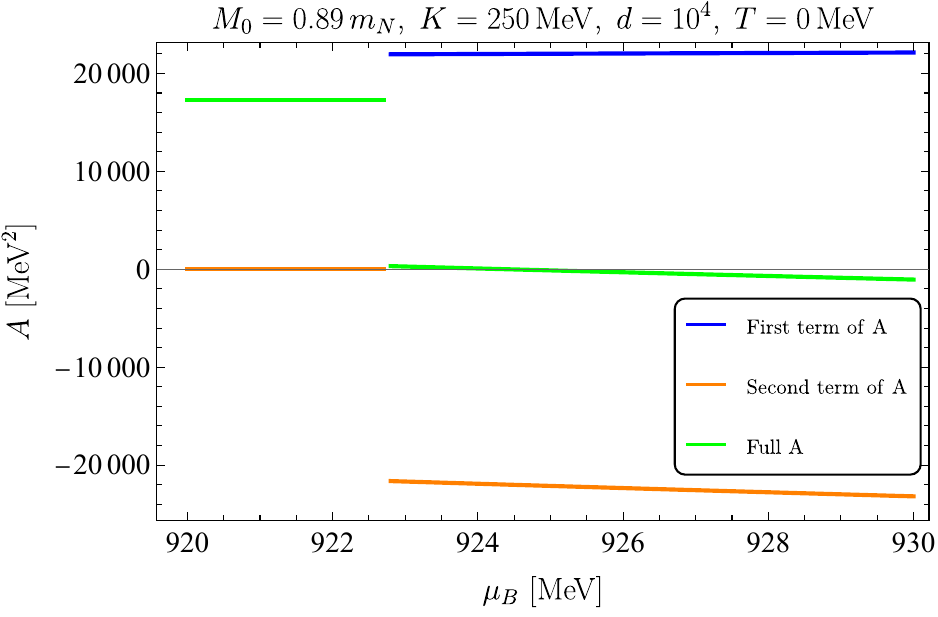}\includegraphics[height=5.6cm]{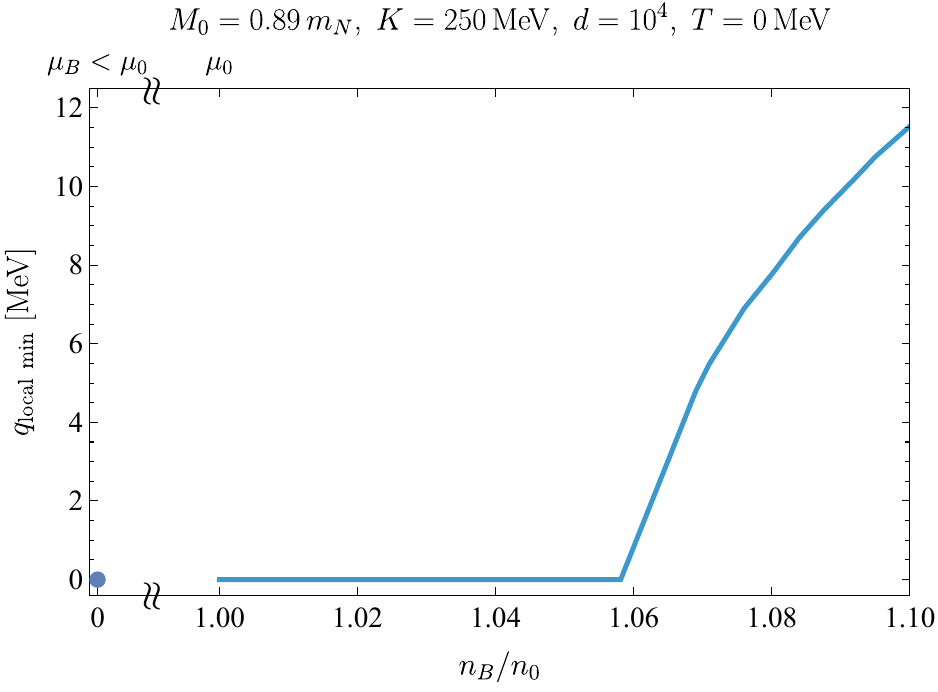}
    \includegraphics[height=5.0cm]{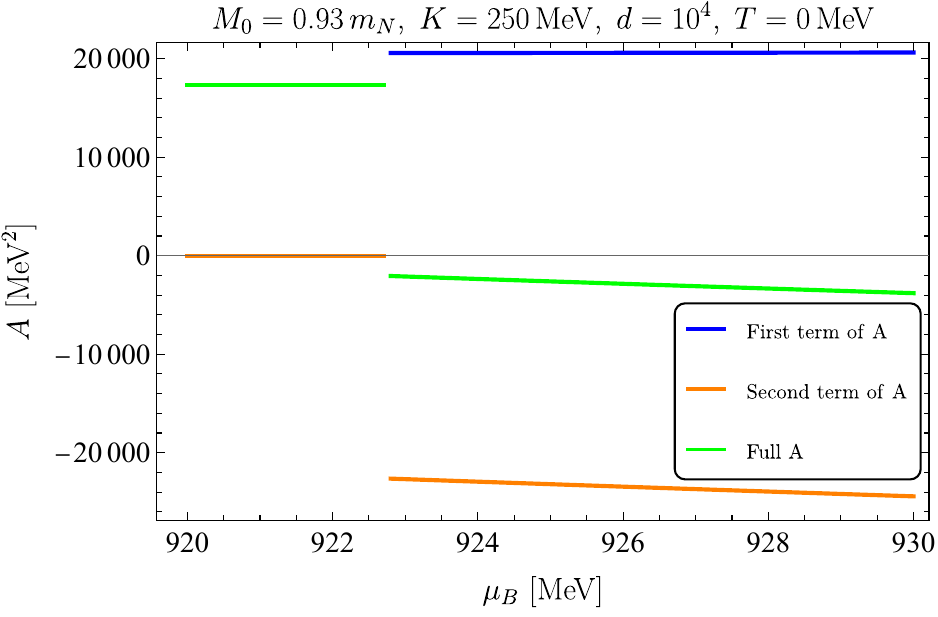}\includegraphics[height=5.6cm]{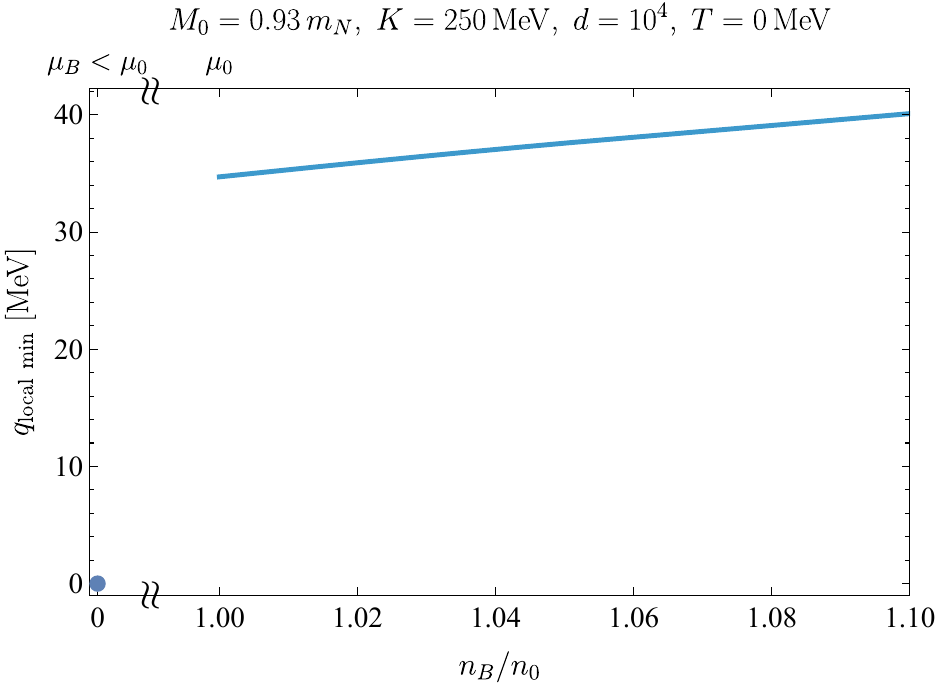}
    \caption{(Left) The dependence of the parameter $A$ and its two terms in  Eq.~\eqref{eq:A-term} on the baryon chemical potential $\mu_B$ at zero temperature for three different values of $M_0$. We fix the other input parameters as $K=250$ MeV, and $d=10^4$. (Right) The wave number $q_{{\rm local}~{\rm min}}$ at the local minimum of $\Omega(q)$ in the vicinity delineated by the left figure.}
    \label{Fig:A-term}
\end{figure*}

Estimating the bubble nucleation rate is highly challenging because we do not know how the order parameter $q$ behaves as a field.
Moreover, there is a subtlety in treating the Kronecker delta term; this should actually be a smooth function connecting from zero to one with a macroscopic length scale, such as the size of the CDW domain.
This scale is completely unknown to us.

Instead, we first check whether the potential barrier can last until a sufficiently low baryon density, which is a necessary condition for the QCD-induced little inflationary scenario.
For instance, in Fig.\,\ref{fig:CDW_phase_diagram}, we have depicted the spinodal line (yellow dashed line) above which  the potential barrier exists. 
As shown in the figure, we find that the CDW phase can only exist at $\mu_B \gtrsim 960\,\MeV$, which is larger than $\mu_0$, implying that the transition ends before the baryon density becomes sufficiently reduced. 
Therefore, the QCD-induced little inflation cannot be realized for the given parameter set of $M_0=0.81\,m_N$, $d=10^4$, and $K=250\,\MeV$.

\begin{figure*}[t!]
    \centering
    \includegraphics[width=8cm]{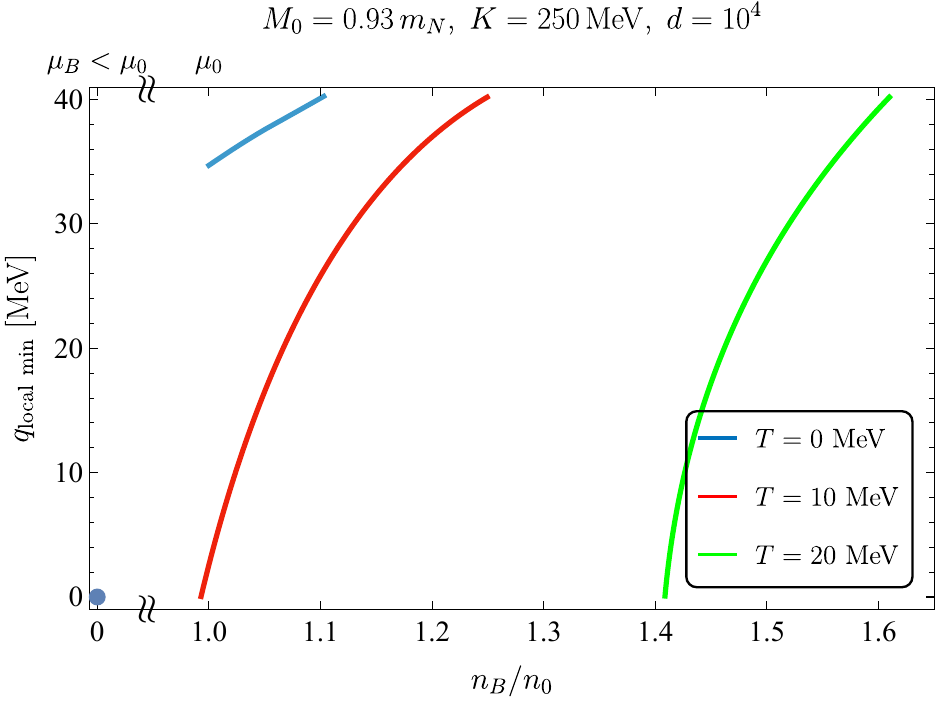}\includegraphics[width=8cm]{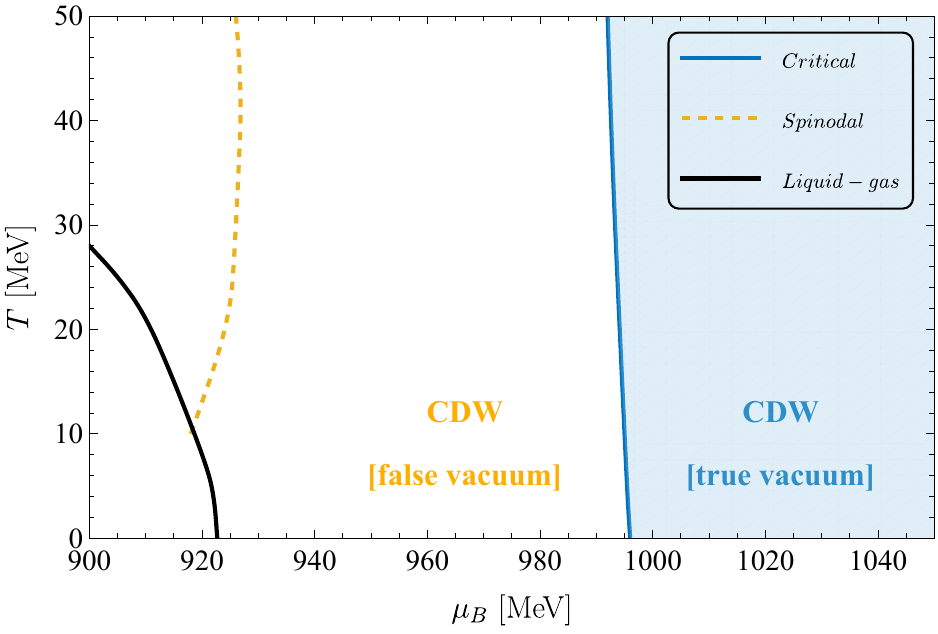}
    \caption{(Left) The wave number $q$ at the local minimum of the thermodynamic potential $\Omega(q)$ as a function of the normalized baryon number density $n_B/n_0$ for three different temperatures, $T=0,\,10,\,20 \ \mathrm{MeV}$. (Right) The corresponding phase diagram in $\mu_B-T$ , for the benchmark parameter $M_0=0.93m_N$, illustrating the regions of the CDW phase as a metastable state (false vacuum) and a stable state (true vacuum). The liquid-gas transition is denoted by a black solid line.}
    \label{fig:CDW_phase_diagram_M0_0.93}
\end{figure*}

In the following, we investigate further by taking a larger value of $M_0$, which makes the CDW phase more stable, e.g., see Fig.\,\ref{fig:CDW_phase_in_M0-muB}. 
We first discuss the case of $T=0$, followed by our main results with $T\neq0$. 
In the former, we are not only able to maximize the (meta)stable CDW phase, but also provide insights on the nature of the spinodal decomposition with a certain level of analytical calculations. 

As discussed in Sec.\,\ref{sec:CDW_phase}, the potential barrier exists as long as there is a local minimum at $q\neq0$, because the Kronecker delta term always makes the homogeneous chiral condensate ($q=0$) local minimum (see Figs.\,\ref{fig:potential_in_q_zero_T} and \ref{fig:potential_in_q_nonzero_T}).
To have a minimum at $q\neq0$, the curvature of the smooth part of $\Delta \Omega$ at $q=0$ should be negative.

Expanding $\Delta \Omega (q)$ around $q=0$, 
we define $A$, the curvature of the potential, as
\bal
\Delta \Omega (q)= (1-\delta_{q0})\epsilon \tilde \phi_0 + A q^2 + {\cal O}(q^4),
\label{eq:omega_expansion}
\eal
and we find
\bal
    A
    = & 2{\tilde \phi_0}^{2}\left(1-\frac{g_{\sigma}^{2}}{4\pi^{2}}\ln\!\frac{M({\tilde \phi_0})^{2}}{m_N^2}\right)
    \nn
    \\
    &- \Theta(\tilde \mu_{*0}-M(\tilde \phi_0))\,\frac{M({\tilde \phi_0})^{2}}{\pi^{2}}
    \ln\!\left(\frac{\mu_{\ast}+k_{F}}{M({\tilde \phi_0})}\right),
    \label{eq:A-term}
\eal
where $M(\tilde \phi_0)= g_\sigma \tilde\phi_0$ with $\tilde \phi_0 \equiv \lim_{q\to0}\phi_{\rm min}(q)$, and $\tilde \mu_{*0}=\mu_B-g_\omega \tilde \omega_0$ with $\tilde \omega_0 = \lim_{q\to0}\omega_{\rm min}(q)$. 
The coefficient of the quartic term $q^4$ is positive, and therefore a local minimum at $q\neq0$ exists only when $A<0$.
Thus, $A=0$ determines the spinodal line.

$A$ in Eq.\,\eqref{eq:A-term} has two contributions.
The first term is always positive because we have a relation of $M(\tilde\phi_0)=(\tilde\phi_0/f_\pi)m_N< m_N$,
which makes $\log (M(\tilde \phi_0)/m_N)$ negative.
The second term is negative, and activated only for the positive value of  $\tilde \mu_{*0}-M(\tilde \phi_0)$, which is equivalent to $\mu_B>\mu_0$, due to the discontinuity in $\phi_{\rm min}$ at $q=0$.
To obtain a negative value of $A$, it is therefore necessary to have a large $\mu_B$, which explains what we have seen in Figs.~\ref{fig:potential_zero_T} and \ref{fig:potential_in_q_zero_T}. 

In the left panels of Fig.\,\ref{Fig:A-term}, we show the first and second terms in Eq.\,\eqref{eq:A-term} by blue and orange lines, respectively, while the whole $A$ term is depicted by the green lines.
From top to bottom, $M_0/m_N$ is taken to increase as $0.81$, $0.89$, and $0.93$. Again, we fix the other input parameters as $K=250\,\MeV$ and $d=10^4$. 
When $\mu_B$ is below $\mu_0$, we find that $A$ is always positive due to the absence of the second term in Eq.\,\eqref{eq:A-term}. 
As shown in the figures, however, for $\mu_B>\mu_0$ the fact that the first and second terms are opposite with comparable magnitudes leads to a small value of $A$. 
We note that both terms scale with $\tilde \phi_0^2$ multiplied by the ${\cal{O}}(1)$ coefficients. 
In particular, the value of $A$ still remains positive just above $\mu_0$ in the cases with $M_0/m_N=0.81$ and $0.89$, but becomes negative for the higher values of $M_0$, as illustrated in the bottom panel for $M_0/m_M=0.93$. 
This indicates that a sufficiently large supercooling of the CDW phase may be possible if $M_0$ is larger than $0.9\,m_N$, where the sign flip of $A$ is expected to occur abruptly at $\mu_B=\mu_0$. 

The location of the \emph{local} minimum at $q\neq 0$ is depicted in the right panels of Fig.\,\ref{Fig:A-term}.
In these plots, we convert $\mu_B$ in terms of $n_B/n_0$, while the upper ticks indicate the $\mu_B$ values corresponding to $n_B/n_0$, e.g. $\mu_0=922.7\,\MeV$ corresponds to $n_B/n_0=1$.
In each panel, as expected, the $q\neq0$ minimum disappears when the $A$ term becomes positive.
More importantly, as discussed above, this happens at $\mu_B=\mu_0$ when $M_0\gtrsim 0.9\,m_N$, and therefore the CDW phase can be supercooled until the baryon onset. 
In other words, the CDW phase can directly transition into the hadronic gas state where $n_B=0$.

Such a behavior still holds for nonzero (low) temperature. 
In the left panel of Fig.\,\ref{fig:CDW_phase_diagram_M0_0.93}, we show the evolution of a local minimum at $q\neq 0$ for three different temperatures, $T=0$ (blue), $10\,\MeV$ (red), and $20\,\MeV$ (green). We find that the local minimum at $q\neq 0$ survives until the liquid-gas transition surface as long as $T\lesssim10\,\MeV$.
The corresponding phase diagram in the $\mu_B$\,--\,$T$ plane is shown in the right panel. 

\subsection{Hadronic liquid-gas transition}
\label{sec:liquid-gas}
To see what happens during the transition between the CDW and hadronic phases for $M_0\gtrsim 0.9\,m_N$, we first need to understand the nature of the hadronic liquid-gas transition.
The baryon number density can be expressed as
\bal
n_B = 4\int \frac{d^3k}{(2\pi)^3} \left[ \frac{1}{e^{(E-\mu_*)/T}+1}-\frac{1}{e^{(E+\mu_*)/T}+1} \right],
\eal
where $E=\sqrt{k^2+M^2}>M$ and  $M=g_\sigma \langle \phi \rangle$.
In the limit of $T\to 0$, if $\mu_*<M$, both exponents of baryon and antibaryon distributions become positive infinity for any $k$.
Thus, $n_B$ drops down to zero discontinuously when $\mu_B < \mu_0$, which leads to the transition from the liquid to the gas state. 
The discontinuity in $n_B$ from $n_0$ to zero across $\mu_0$ suggests that the liquid-gas transition is first order.

\begin{figure}[t]
    \centering
    \includegraphics[width=0.48\textwidth]{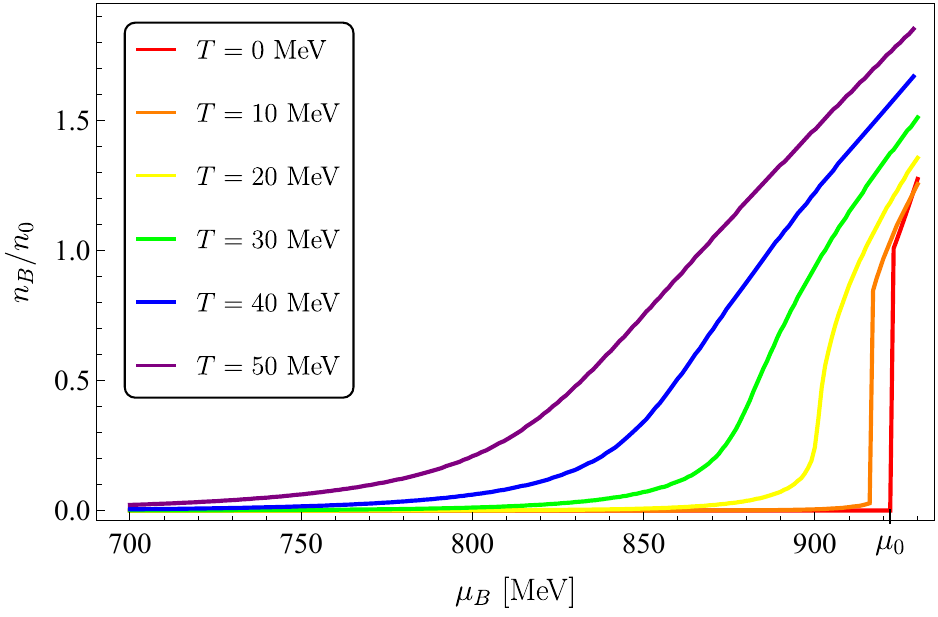}
    \caption{
    Baryon number density as a function of the chemical potential for different temperatures. 
    The input parameters are fixed by $M_0=0.81\,m_N$, $K=250\,\MeV$ and $d=10^4$}
    \label{fig:M_and_nB}
\end{figure}

The liquid-gas transition keeps being first order even when we consider a small, nonzero temperature.
In Fig.\,\ref{fig:M_and_nB}, we depict $n_B/n_0$ as a function of $\mu_B$ for different temperatures.
As shown in the figure, the transition is first-order when $T\lesssim 20\,\MeV$, and becomes a crossover at higher temperature.

It is interesting to consider how the liquid-gas transition proceeds cosmologically, assuming that it exists.
Let us imagine $T=0$ for simplicity.
In the cosmological aspect, since the Universe expands, the baryon number density scales as $a^{-3}$ \emph{continuously}, denoting $a$ as the scale factor.
Thus, at some point, the baryon density must be smaller than $n_0$.
However, Fig.\,\ref{fig:M_and_nB} suggests that there is no chemical potential that can describe $n_B$ below $n_0$.
Taking a small but nonzero $T$ does not help as long as there is a range of $n_B$ in which the baryon chemical potential is not thermodynamically well defined. 

This apparent contradiction can be resolved by the formation of hadronic liquid droplets (e.g. see Ref.\,\cite{1999nap..book.6288N}), breaking the homogeneity spontaneously.
Each droplet maintains the baryon number density $n_B=n_0$.
The space between droplets is filled by the hadronic gas state ($n_B=0$ for $T=0$), while the distances among droplets get farther proportionally to $a$.
Therefore, the \emph{net} baryon density can be scaled as $a^{-3}$.

\subsection{Cosmological implication}

Let us come back to the CDW phase with $M_0 \gtrsim 0.9\,m_N$, which is the only remaining setup that may provide a strong supercooling of the CDW phase, putting aside the fact that such a large $M_0$ value is disfavored.
We showed that the CDW phase can last until the liquid-gas transition surface, and we now need to consider a transition from the CDW phase to the hadronic gas state.

Recall that, as described in Sec.\,\ref{sec:liquid-gas}, the hadronic liquid-gas transition proceeds with the formation of liquid droplets whose density is $n_0$.
The same picture must be applied here---when the net baryon density touches the saturation density, droplets form, while the space between them is filled by the hadronic gas state.
As the Universe expands, the distance among droplets scales linearly in $a$, and the net baryon density decreases as $a^{-3}$.

Each droplet can either stay in the CDW phase if the size of the droplet is sufficiently large, or be forced to transition into the homogeneous chiral condensate.
When the CDW phase makes the transition inside a droplet, the potential energy difference between $q\neq0$ and $q=0$ minima is released as latent heat.
Assuming the simultaneous transition and instantaneous thermalization of the released heat, the reheating temperature $T_{\rm RH}$ after the transition can be estimated as
\bal
\rho_{\rm pl}(T_{\rm RH}) = \Delta \Omega(\tilde q_0) V_{\rm d} n_{\rm d},
\label{eq:T_RH}
\eal
where $\rho_{\rm pl}$ is the plasma energy density, $V_{\rm d}$ is the volume of each droplet, $n_{\rm d}$ is the number density of droplets, and $\tilde q_0$ is the wave number at the local minimum in the limit of $\mu_B\to \mu_0+$.
Since the potential energy difference mostly comes from the Kronecker delta term in Eq.\,\eqref{eq:Delta_U}, we approximate $\Delta \Omega(\tilde q_0)\simeq \epsilon f_\pi =m_\pi^2 f_\pi^2$, which is actually an optimistic estimate for the upper bound of $\Delta \Omega(\tilde q_0)$. 

The net baryon density becomes
\bal
n_B^{\rm net} = n_0 V_{\rm d} n_{\rm d} = n_0 \frac{\rho_{\rm pl}(T_{\rm RH})}{\Delta \Omega(\tilde q_0)},
\eal
and therefore, taking the entropy density $s=\frac{4}{3}\frac{\rho_{\rm pl}}{T}$, the final baryon yield after the transition can be estimated as
\bal
Y_B &= \frac{n_B^{\rm net}}{s}
=\frac{3}{4} \frac{n_0 T_{\rm RH}}{\Delta \Omega(\tilde q_0)}
\\
&\simeq
10^{-2} \frac{T_{\rm RH}}{2\,\MeV}.
\eal
As $Y_B\sim 10^{-10}$ requires the reheating temperature to be significantly lower than the neutrino decoupling temperature $\sim 2\,\MeV$, this scenario is severely ruled out by BBN and CMB.

We thus conclude that the little inflationary scenario associated with the CDW phase transition is incompatible with the observed Universe.
Although this conclusion has been derived within a specific nucleon-meson model, we do expect that the qualitative conclusion would not be substantially altered in other models. 
The suppression in Eq.\,\eqref{eq:T_RH} by the factor $V_{\rm d} n_{\rm d}<1$ follows from the nature of the liquid-gas transition which is model independent. 
A different model may slightly increase $\Delta \Omega(\tilde q_0)$, but not by several orders of magnitude.

\section{Summary}
\label{sec:summary}
In this work, we have revisited the QCD-induced little inflationary scenario, which may explain the GW signal observed from PTA experiments.
We mainly focused on the maximally allowed strength of the supercooling in first-order QCD phase transitions, and investigated whether any of such phase transitions can sufficiently dilute the initial large baryon density to be consistent with the  observation of the current Universe. 

We first pointed out that the originally suggested scenario, based on the transition from the QGP to the hadronic phase, cannot realize this scenario.
This is because the existence of the CEP at $\mu_B/T > O(1)$ conflicts with the assumption that a potential barrier separating two phases lasts until a very low temperature and chemical potential.
For instance, the dilaton-quark-meson model studied in the original work predicts a first-order transition even at $\mu_B=0$ and thus is unrealistic.

Then we considered the transition between the inhomogeneous CDW phase and the homogeneous hadronic phase with a nonzero chiral condensate.
Although the existence of the CDW phase in QCD matter has not yet been fully established and still remains a theoretical possibility, we find that, in the nucleon-meson model, its phase transition can be strongly first order when the saturation mass $M_0$ is large and the self-interaction of the $\omega$ meson is strong.
Especially, when $M_0 \gtrsim 0.9\,m_N$ and $d\sim 10^4$, the potential barrier separating $q=0$ and $q\neq 0$ can last until the liquid-gas transition surface, making a large dilution of baryon density possible.
However, the net latent heat is suppressed by the nature of the liquid-gas transition, and thus, the reheating temperature turns out to be as small as the electronvolt scale to successfully explain the observed baryon yield of the current Universe.
Such a low reheating temperature is incompatible with the BBN and CMB, and thus it closes the possibility of realizing the QCD-induced little inflationary scenario and explaining the GW signal observed at PTAs.
We do not expect that this conclusion would change in a different model because the crucial difficulty comes from the model-independent nature of the liquid-gas transition that is accompanied by the strongly supercooled CDW/hadronic phase transition.
It is still conceivable that a color-superconducting phase at much higher baryon density could undergo sufficiently strong supercooling, though we consider this unlikely.

Introducing an additional dilution mechanism, such as early matter domination, may help dilution of the baryon density to be consistent with the observation.
In this case, the GWs produced by a first-order QCD phase transition must also be weakened by the same dilution mechanism, predicting the GW signal to be hidden behind the GW observed at PTAs.

In short, we conclude that it is difficult to realize the QCD-induced little inflationary scenario within standard model QCD, and that a first-order QCD phase transition is therefore unlikely to account for the observed stochastic GW signal. 
More general scenarios involving additional dynamics in the QCD era may nevertheless lead to qualitatively different possibilities, which deserve further investigation.

\section*{Acknowledgments}
We would like to thank Seung-il Nam for the useful comments on the effective treatment of the $\omega$ meson self-interaction. S.\,K. is supported by the National Research Foundation (NRF) under Grant No. RS-2008-NR007226 and by the Institute of Information and Communication Technology Planning and Evaluation grant IITP-2024-RS-2024-00437191 funded by the Korean government (Ministry of Science and ICT).
The work of T.\,H.\,J., J.-W.\,L. and C.\,S.\,S. was supported by IBS under the project code IBS-R018-D1. C.\,S.\,S. and H.\,B.\,Y. are supported by Global-Learning \& Academic research institution for Master’s, PhD students, and Postdocs (G- LAMP) Program of the National Research Foundation of Korea (NRF) grants funded by the Ministry of Education (No. RS-2025025442707), and by the NRF grants funded by the Korean government (MSIT) (No. RS-2026- 25498521).

\bibliography{ref.bib}

\begin{thebibliography}{65}%
\makeatletter
\providecommand \@ifxundefined [1]{%
 \@ifx{#1\undefined}
}%
\providecommand \@ifnum [1]{%
 \ifnum #1\expandafter \@firstoftwo
 \else \expandafter \@secondoftwo
 \fi
}%
\providecommand \@ifx [1]{%
 \ifx #1\expandafter \@firstoftwo
 \else \expandafter \@secondoftwo
 \fi
}%
\providecommand \natexlab [1]{#1}%
\providecommand \enquote  [1]{``#1''}%
\providecommand \bibnamefont  [1]{#1}%
\providecommand \bibfnamefont [1]{#1}%
\providecommand \citenamefont [1]{#1}%
\providecommand \href@noop [0]{\@secondoftwo}%
\providecommand \href [0]{\begingroup \@sanitize@url \@href}%
\providecommand \@href[1]{\@@startlink{#1}\@@href}%
\providecommand \@@href[1]{\endgroup#1\@@endlink}%
\providecommand \@sanitize@url [0]{\catcode `\\12\catcode `\$12\catcode
  `\&12\catcode `\#12\catcode `\^12\catcode `\_12\catcode `\%12\relax}%
\providecommand \@@startlink[1]{}%
\providecommand \@@endlink[0]{}%
\providecommand \url  [0]{\begingroup\@sanitize@url \@url }%
\providecommand \@url [1]{\endgroup\@href {#1}{\urlprefix }}%
\providecommand \urlprefix  [0]{URL }%
\providecommand \Eprint [0]{\href }%
\providecommand \doibase [0]{https://doi.org/}%
\providecommand \selectlanguage [0]{\@gobble}%
\providecommand \bibinfo  [0]{\@secondoftwo}%
\providecommand \bibfield  [0]{\@secondoftwo}%
\providecommand \translation [1]{[#1]}%
\providecommand \BibitemOpen [0]{}%
\providecommand \bibitemStop [0]{}%
\providecommand \bibitemNoStop [0]{.\EOS\space}%
\providecommand \EOS [0]{\spacefactor3000\relax}%
\providecommand \BibitemShut  [1]{\csname bibitem#1\endcsname}%
\let\auto@bib@innerbib\@empty
\bibitem [{\citenamefont {Agazie}\ \emph {et~al.}(2023)\citenamefont {Agazie}
  \emph {et~al.}}]{NANOGrav:2023gor}%
  \BibitemOpen
  \bibfield  {author} {\bibinfo {author} {\bibfnamefont {G.}~\bibnamefont
  {Agazie}} \emph {et~al.} (\bibinfo {collaboration} {NANOGrav}),\ }\bibfield
  {title} {\bibinfo {title} {{The NANOGrav 15 yr Data Set: Evidence for a
  Gravitational-wave Background}},\ }\href
  {https://doi.org/10.3847/2041-8213/acdac6} {\bibfield  {journal} {\bibinfo
  {journal} {Astrophys. J. Lett.}\ }\textbf {\bibinfo {volume} {951}},\
  \bibinfo {pages} {L8} (\bibinfo {year} {2023})},\ \Eprint
  {https://arxiv.org/abs/2306.16213} {arXiv:2306.16213 [astro-ph.HE]}
  \BibitemShut {NoStop}%
\bibitem [{\citenamefont {Antoniadis}\ \emph {et~al.}(2023)\citenamefont
  {Antoniadis} \emph {et~al.}}]{EPTA:2023fyk}%
  \BibitemOpen
  \bibfield  {author} {\bibinfo {author} {\bibfnamefont {J.}~\bibnamefont
  {Antoniadis}} \emph {et~al.} (\bibinfo {collaboration} {EPTA, InPTA:}),\
  }\bibfield  {title} {\bibinfo {title} {{The second data release from the
  European Pulsar Timing Array - III. Search for gravitational wave signals}},\
  }\href {https://doi.org/10.1051/0004-6361/202346844} {\bibfield  {journal}
  {\bibinfo  {journal} {Astron. Astrophys.}\ }\textbf {\bibinfo {volume}
  {678}},\ \bibinfo {pages} {A50} (\bibinfo {year} {2023})},\ \Eprint
  {https://arxiv.org/abs/2306.16214} {arXiv:2306.16214 [astro-ph.HE]}
  \BibitemShut {NoStop}%
\bibitem [{\citenamefont {Reardon}\ \emph {et~al.}(2023)\citenamefont {Reardon}
  \emph {et~al.}}]{Reardon:2023gzh}%
  \BibitemOpen
  \bibfield  {author} {\bibinfo {author} {\bibfnamefont {D.~J.}\ \bibnamefont
  {Reardon}} \emph {et~al.},\ }\bibfield  {title} {\bibinfo {title} {{Search
  for an Isotropic Gravitational-wave Background with the Parkes Pulsar Timing
  Array}},\ }\href {https://doi.org/10.3847/2041-8213/acdd02} {\bibfield
  {journal} {\bibinfo  {journal} {Astrophys. J. Lett.}\ }\textbf {\bibinfo
  {volume} {951}},\ \bibinfo {pages} {L6} (\bibinfo {year} {2023})},\ \Eprint
  {https://arxiv.org/abs/2306.16215} {arXiv:2306.16215 [astro-ph.HE]}
  \BibitemShut {NoStop}%
\bibitem [{\citenamefont {Xu}\ \emph {et~al.}(2023)\citenamefont {Xu} \emph
  {et~al.}}]{Xu:2023wog}%
  \BibitemOpen
  \bibfield  {author} {\bibinfo {author} {\bibfnamefont {H.}~\bibnamefont {Xu}}
  \emph {et~al.},\ }\bibfield  {title} {\bibinfo {title} {{Searching for the
  Nano-Hertz Stochastic Gravitational Wave Background with the Chinese Pulsar
  Timing Array Data Release I}},\ }\href
  {https://doi.org/10.1088/1674-4527/acdfa5} {\bibfield  {journal} {\bibinfo
  {journal} {Res. Astron. Astrophys.}\ }\textbf {\bibinfo {volume} {23}},\
  \bibinfo {pages} {075024} (\bibinfo {year} {2023})},\ \Eprint
  {https://arxiv.org/abs/2306.16216} {arXiv:2306.16216 [astro-ph.HE]}
  \BibitemShut {NoStop}%
\bibitem [{\citenamefont {Afzal}\ \emph {et~al.}(2023)\citenamefont {Afzal}
  \emph {et~al.}}]{NANOGrav:2023hvm}%
  \BibitemOpen
  \bibfield  {author} {\bibinfo {author} {\bibfnamefont {A.}~\bibnamefont
  {Afzal}} \emph {et~al.} (\bibinfo {collaboration} {NANOGrav}),\ }\bibfield
  {title} {\bibinfo {title} {{The NANOGrav 15 yr Data Set: Search for Signals
  from New Physics}},\ }\href {https://doi.org/10.3847/2041-8213/acdc91}
  {\bibfield  {journal} {\bibinfo  {journal} {Astrophys. J. Lett.}\ }\textbf
  {\bibinfo {volume} {951}},\ \bibinfo {pages} {L11} (\bibinfo {year}
  {2023})},\ \bibinfo {note} {[Erratum: Astrophys.J.Lett. 971, L27 (2024),
  Erratum: Astrophys.J. 971, L27 (2024)]},\ \Eprint
  {https://arxiv.org/abs/2306.16219} {arXiv:2306.16219 [astro-ph.HE]}
  \BibitemShut {NoStop}%
\bibitem [{\citenamefont {Witten}(1984)}]{Witten:1984rs}%
  \BibitemOpen
  \bibfield  {author} {\bibinfo {author} {\bibfnamefont {E.}~\bibnamefont
  {Witten}},\ }\bibfield  {title} {\bibinfo {title} {{Cosmic Separation of
  Phases}},\ }\href {https://doi.org/10.1103/PhysRevD.30.272} {\bibfield
  {journal} {\bibinfo  {journal} {Phys. Rev. D}\ }\textbf {\bibinfo {volume}
  {30}},\ \bibinfo {pages} {272} (\bibinfo {year} {1984})}\BibitemShut
  {NoStop}%
\bibitem [{\citenamefont {Boeckel}\ and\ \citenamefont
  {Schaffner-Bielich}(2010)}]{Boeckel:2009ej}%
  \BibitemOpen
  \bibfield  {author} {\bibinfo {author} {\bibfnamefont {T.}~\bibnamefont
  {Boeckel}}\ and\ \bibinfo {author} {\bibfnamefont {J.}~\bibnamefont
  {Schaffner-Bielich}},\ }\bibfield  {title} {\bibinfo {title} {{A little
  inflation in the early universe at the QCD phase transition}},\ }\href
  {https://doi.org/10.1103/PhysRevLett.105.041301} {\bibfield  {journal}
  {\bibinfo  {journal} {Phys. Rev. Lett.}\ }\textbf {\bibinfo {volume} {105}},\
  \bibinfo {pages} {041301} (\bibinfo {year} {2010})},\ \bibinfo {note}
  {[Erratum: Phys.Rev.Lett. 106, 069901 (2011)]},\ \Eprint
  {https://arxiv.org/abs/0906.4520} {arXiv:0906.4520 [astro-ph.CO]}
  \BibitemShut {NoStop}%
\bibitem [{\citenamefont {Boeckel}\ and\ \citenamefont
  {Schaffner-Bielich}(2012)}]{Boeckel:2011yj}%
  \BibitemOpen
  \bibfield  {author} {\bibinfo {author} {\bibfnamefont {T.}~\bibnamefont
  {Boeckel}}\ and\ \bibinfo {author} {\bibfnamefont {J.}~\bibnamefont
  {Schaffner-Bielich}},\ }\bibfield  {title} {\bibinfo {title} {{A little
  inflation at the cosmological QCD phase transition}},\ }\href
  {https://doi.org/10.1103/PhysRevD.85.103506} {\bibfield  {journal} {\bibinfo
  {journal} {Phys. Rev. D}\ }\textbf {\bibinfo {volume} {85}},\ \bibinfo
  {pages} {103506} (\bibinfo {year} {2012})},\ \Eprint
  {https://arxiv.org/abs/1105.0832} {arXiv:1105.0832 [astro-ph.CO]}
  \BibitemShut {NoStop}%
\bibitem [{\citenamefont {Schettler}\ \emph {et~al.}(2011)\citenamefont
  {Schettler}, \citenamefont {Boeckel},\ and\ \citenamefont
  {Schaffner-Bielich}}]{Schettler:2010dp}%
  \BibitemOpen
  \bibfield  {author} {\bibinfo {author} {\bibfnamefont {S.}~\bibnamefont
  {Schettler}}, \bibinfo {author} {\bibfnamefont {T.}~\bibnamefont {Boeckel}},\
  and\ \bibinfo {author} {\bibfnamefont {J.}~\bibnamefont
  {Schaffner-Bielich}},\ }\bibfield  {title} {\bibinfo {title} {{Imprints of
  the QCD Phase Transition on the Spectrum of Gravitational Waves}},\ }\href
  {https://doi.org/10.1103/PhysRevD.83.064030} {\bibfield  {journal} {\bibinfo
  {journal} {Phys. Rev. D}\ }\textbf {\bibinfo {volume} {83}},\ \bibinfo
  {pages} {064030} (\bibinfo {year} {2011})},\ \Eprint
  {https://arxiv.org/abs/1010.4857} {arXiv:1010.4857 [astro-ph.CO]}
  \BibitemShut {NoStop}%
\bibitem [{\citenamefont {McInnes}(2016)}]{McInnes:2015hga}%
  \BibitemOpen
  \bibfield  {author} {\bibinfo {author} {\bibfnamefont {B.}~\bibnamefont
  {McInnes}},\ }\bibfield  {title} {\bibinfo {title} {{Trajectory of the cosmic
  plasma through the quark matter phase diagram}},\ }\href
  {https://doi.org/10.1103/PhysRevD.93.043544} {\bibfield  {journal} {\bibinfo
  {journal} {Phys. Rev. D}\ }\textbf {\bibinfo {volume} {93}},\ \bibinfo
  {pages} {043544} (\bibinfo {year} {2016})},\ \Eprint
  {https://arxiv.org/abs/1506.05873} {arXiv:1506.05873 [hep-th]} \BibitemShut
  {NoStop}%
\bibitem [{\citenamefont {Ahmadvand}\ and\ \citenamefont
  {Bitaghsir~Fadafan}(2017)}]{Ahmadvand:2017xrw}%
  \BibitemOpen
  \bibfield  {author} {\bibinfo {author} {\bibfnamefont {M.}~\bibnamefont
  {Ahmadvand}}\ and\ \bibinfo {author} {\bibfnamefont {K.}~\bibnamefont
  {Bitaghsir~Fadafan}},\ }\bibfield  {title} {\bibinfo {title} {{Gravitational
  waves generated from the cosmological QCD phase transition within AdS/QCD}},\
  }\href {https://doi.org/10.1016/j.physletb.2017.07.039} {\bibfield  {journal}
  {\bibinfo  {journal} {Phys. Lett. B}\ }\textbf {\bibinfo {volume} {772}},\
  \bibinfo {pages} {747} (\bibinfo {year} {2017})},\ \Eprint
  {https://arxiv.org/abs/1703.02801} {arXiv:1703.02801 [hep-th]} \BibitemShut
  {NoStop}%
\bibitem [{\citenamefont {He}\ \emph {et~al.}(2025)\citenamefont {He},
  \citenamefont {Li}, \citenamefont {Wang},\ and\ \citenamefont
  {Wang}}]{He:2023ado}%
  \BibitemOpen
  \bibfield  {author} {\bibinfo {author} {\bibfnamefont {S.}~\bibnamefont
  {He}}, \bibinfo {author} {\bibfnamefont {L.}~\bibnamefont {Li}}, \bibinfo
  {author} {\bibfnamefont {S.}~\bibnamefont {Wang}},\ and\ \bibinfo {author}
  {\bibfnamefont {S.-J.}\ \bibnamefont {Wang}},\ }\bibfield  {title} {\bibinfo
  {title} {{Constraints on holographic QCD phase transitions from PTA
  observations}},\ }\href {https://doi.org/10.1007/s11433-024-2468-x}
  {\bibfield  {journal} {\bibinfo  {journal} {Sci. China Phys. Mech. Astron.}\
  }\textbf {\bibinfo {volume} {68}},\ \bibinfo {pages} {210411} (\bibinfo
  {year} {2025})},\ \Eprint {https://arxiv.org/abs/2308.07257}
  {arXiv:2308.07257 [hep-ph]} \BibitemShut {NoStop}%
\bibitem [{\citenamefont {Han}\ and\ \citenamefont {Shao}(2023)}]{Han:2023znh}%
  \BibitemOpen
  \bibfield  {author} {\bibinfo {author} {\bibfnamefont {X.}~\bibnamefont
  {Han}}\ and\ \bibinfo {author} {\bibfnamefont {G.}~\bibnamefont {Shao}},\
  }\bibfield  {title} {\bibinfo {title} {{Stochastic gravitational waves
  produced by the first-order QCD phase transition}},\ }\href@noop {} {\
  (\bibinfo {year} {2023})},\ \Eprint {https://arxiv.org/abs/2312.00571}
  {arXiv:2312.00571 [astro-ph.CO]} \BibitemShut {NoStop}%
\bibitem [{\citenamefont {Shao}\ \emph
  {et~al.}(2025{\natexlab{a}})\citenamefont {Shao}, \citenamefont {Mao},\ and\
  \citenamefont {Huang}}]{Shao:2024ygm}%
  \BibitemOpen
  \bibfield  {author} {\bibinfo {author} {\bibfnamefont {J.}~\bibnamefont
  {Shao}}, \bibinfo {author} {\bibfnamefont {H.}~\bibnamefont {Mao}},\ and\
  \bibinfo {author} {\bibfnamefont {M.}~\bibnamefont {Huang}},\ }\bibfield
  {title} {\bibinfo {title} {{Nanohertz gravitational waves and primordial
  quark nuggets from dense QCD matter in the early Universe*}},\ }\href
  {https://doi.org/10.1088/1674-1137/adbeeb} {\bibfield  {journal} {\bibinfo
  {journal} {Chin. Phys. C}\ }\textbf {\bibinfo {volume} {49}},\ \bibinfo
  {pages} {065103} (\bibinfo {year} {2025}{\natexlab{a}})},\ \Eprint
  {https://arxiv.org/abs/2410.00874} {arXiv:2410.00874 [hep-ph]} \BibitemShut
  {NoStop}%
\bibitem [{\citenamefont {Shao}\ \emph
  {et~al.}(2025{\natexlab{b}})\citenamefont {Shao}, \citenamefont {Mao},\ and\
  \citenamefont {Huang}}]{Shao:2024dxt}%
  \BibitemOpen
  \bibfield  {author} {\bibinfo {author} {\bibfnamefont {J.}~\bibnamefont
  {Shao}}, \bibinfo {author} {\bibfnamefont {H.}~\bibnamefont {Mao}},\ and\
  \bibinfo {author} {\bibfnamefont {M.}~\bibnamefont {Huang}},\ }\bibfield
  {title} {\bibinfo {title} {{Transition rate and gravitational wave spectrum
  from first-order QCD phase transitions}},\ }\href
  {https://doi.org/10.1103/PhysRevD.111.023052} {\bibfield  {journal} {\bibinfo
   {journal} {Phys. Rev. D}\ }\textbf {\bibinfo {volume} {111}},\ \bibinfo
  {pages} {023052} (\bibinfo {year} {2025}{\natexlab{b}})},\ \Eprint
  {https://arxiv.org/abs/2410.06780} {arXiv:2410.06780 [hep-ph]} \BibitemShut
  {NoStop}%
\bibitem [{\citenamefont {Schwarz}\ and\ \citenamefont
  {Stuke}(2009)}]{Schwarz:2009ii}%
  \BibitemOpen
  \bibfield  {author} {\bibinfo {author} {\bibfnamefont {D.~J.}\ \bibnamefont
  {Schwarz}}\ and\ \bibinfo {author} {\bibfnamefont {M.}~\bibnamefont
  {Stuke}},\ }\bibfield  {title} {\bibinfo {title} {{Lepton asymmetry and the
  cosmic QCD transition}},\ }\href
  {https://doi.org/10.1088/1475-7516/2009/11/025} {\bibfield  {journal}
  {\bibinfo  {journal} {JCAP}\ }\textbf {\bibinfo {volume} {11}},\ \bibinfo
  {pages} {025}},\ \bibinfo {note} {[Erratum: JCAP 10, E01 (2010)]},\ \Eprint
  {https://arxiv.org/abs/0906.3434} {arXiv:0906.3434 [hep-ph]} \BibitemShut
  {NoStop}%
\bibitem [{\citenamefont {Caprini}\ \emph {et~al.}(2010)\citenamefont
  {Caprini}, \citenamefont {Durrer},\ and\ \citenamefont
  {Siemens}}]{Caprini:2010xv}%
  \BibitemOpen
  \bibfield  {author} {\bibinfo {author} {\bibfnamefont {C.}~\bibnamefont
  {Caprini}}, \bibinfo {author} {\bibfnamefont {R.}~\bibnamefont {Durrer}},\
  and\ \bibinfo {author} {\bibfnamefont {X.}~\bibnamefont {Siemens}},\
  }\bibfield  {title} {\bibinfo {title} {{Detection of gravitational waves from
  the QCD phase transition with pulsar timing arrays}},\ }\href
  {https://doi.org/10.1103/PhysRevD.82.063511} {\bibfield  {journal} {\bibinfo
  {journal} {Phys. Rev. D}\ }\textbf {\bibinfo {volume} {82}},\ \bibinfo
  {pages} {063511} (\bibinfo {year} {2010})},\ \Eprint
  {https://arxiv.org/abs/1007.1218} {arXiv:1007.1218 [astro-ph.CO]}
  \BibitemShut {NoStop}%
\bibitem [{\citenamefont {Wygas}\ \emph {et~al.}(2018)\citenamefont {Wygas},
  \citenamefont {Oldengott}, \citenamefont {B{\"o}deker},\ and\ \citenamefont
  {Schwarz}}]{Wygas:2018otj}%
  \BibitemOpen
  \bibfield  {author} {\bibinfo {author} {\bibfnamefont {M.~M.}\ \bibnamefont
  {Wygas}}, \bibinfo {author} {\bibfnamefont {I.~M.}\ \bibnamefont
  {Oldengott}}, \bibinfo {author} {\bibfnamefont {D.}~\bibnamefont
  {B{\"o}deker}},\ and\ \bibinfo {author} {\bibfnamefont {D.~J.}\ \bibnamefont
  {Schwarz}},\ }\bibfield  {title} {\bibinfo {title} {{Cosmic QCD Epoch at
  Nonvanishing Lepton Asymmetry}},\ }\href
  {https://doi.org/10.1103/PhysRevLett.121.201302} {\bibfield  {journal}
  {\bibinfo  {journal} {Phys. Rev. Lett.}\ }\textbf {\bibinfo {volume} {121}},\
  \bibinfo {pages} {201302} (\bibinfo {year} {2018})},\ \Eprint
  {https://arxiv.org/abs/1807.10815} {arXiv:1807.10815 [hep-ph]} \BibitemShut
  {NoStop}%
\bibitem [{\citenamefont {Gao}\ and\ \citenamefont
  {Oldengott}(2022)}]{Gao:2021nwz}%
  \BibitemOpen
  \bibfield  {author} {\bibinfo {author} {\bibfnamefont {F.}~\bibnamefont
  {Gao}}\ and\ \bibinfo {author} {\bibfnamefont {I.~M.}\ \bibnamefont
  {Oldengott}},\ }\bibfield  {title} {\bibinfo {title} {{Cosmology Meets
  Functional QCD: First-Order Cosmic QCD Transition Induced by Large Lepton
  Asymmetries}},\ }\href {https://doi.org/10.1103/PhysRevLett.128.131301}
  {\bibfield  {journal} {\bibinfo  {journal} {Phys. Rev. Lett.}\ }\textbf
  {\bibinfo {volume} {128}},\ \bibinfo {pages} {131301} (\bibinfo {year}
  {2022})},\ \Eprint {https://arxiv.org/abs/2106.11991} {arXiv:2106.11991
  [hep-ph]} \BibitemShut {NoStop}%
\bibitem [{\citenamefont {Gao}\ \emph {et~al.}(2025)\citenamefont {Gao},
  \citenamefont {Harz}, \citenamefont {Hati}, \citenamefont {Lu}, \citenamefont
  {Oldengott},\ and\ \citenamefont {White}}]{Gao:2024fhm}%
  \BibitemOpen
  \bibfield  {author} {\bibinfo {author} {\bibfnamefont {F.}~\bibnamefont
  {Gao}}, \bibinfo {author} {\bibfnamefont {J.}~\bibnamefont {Harz}}, \bibinfo
  {author} {\bibfnamefont {C.}~\bibnamefont {Hati}}, \bibinfo {author}
  {\bibfnamefont {Y.}~\bibnamefont {Lu}}, \bibinfo {author} {\bibfnamefont
  {I.~M.}\ \bibnamefont {Oldengott}},\ and\ \bibinfo {author} {\bibfnamefont
  {G.}~\bibnamefont {White}},\ }\bibfield  {title} {\bibinfo {title}
  {{Baryogenesis and first-order QCD transition with gravitational waves from a
  large lepton asymmetry}},\ }\href {https://doi.org/10.1007/JHEP06(2025)247}
  {\bibfield  {journal} {\bibinfo  {journal} {JHEP}\ }\textbf {\bibinfo
  {volume} {06}},\ \bibinfo {pages} {247}},\ \Eprint
  {https://arxiv.org/abs/2407.17549} {arXiv:2407.17549 [hep-ph]} \BibitemShut
  {NoStop}%
\bibitem [{\citenamefont {Aoki}\ \emph {et~al.}(2006)\citenamefont {Aoki},
  \citenamefont {Fodor}, \citenamefont {Katz},\ and\ \citenamefont
  {Szabo}}]{Aoki:2006br}%
  \BibitemOpen
  \bibfield  {author} {\bibinfo {author} {\bibfnamefont {Y.}~\bibnamefont
  {Aoki}}, \bibinfo {author} {\bibfnamefont {Z.}~\bibnamefont {Fodor}},
  \bibinfo {author} {\bibfnamefont {S.~D.}\ \bibnamefont {Katz}},\ and\
  \bibinfo {author} {\bibfnamefont {K.~K.}\ \bibnamefont {Szabo}},\ }\bibfield
  {title} {\bibinfo {title} {{The QCD transition temperature: Results with
  physical masses in the continuum limit}},\ }\href
  {https://doi.org/10.1016/j.physletb.2006.10.021} {\bibfield  {journal}
  {\bibinfo  {journal} {Phys. Lett. B}\ }\textbf {\bibinfo {volume} {643}},\
  \bibinfo {pages} {46} (\bibinfo {year} {2006})},\ \Eprint
  {https://arxiv.org/abs/hep-lat/0609068} {arXiv:hep-lat/0609068} \BibitemShut
  {NoStop}%
\bibitem [{\citenamefont {Kämpfer}(1986)}]{kampfer:1986}%
  \BibitemOpen
  \bibfield  {author} {\bibinfo {author} {\bibfnamefont {B.}~\bibnamefont
  {Kämpfer}},\ }\bibfield  {title} {\bibinfo {title} {Entropy production
  during an isothermal phase transition in the early universe},\ }\href
  {https://doi.org/https://doi.org/10.1002/asna.2113070403} {\bibfield
  {journal} {\bibinfo  {journal} {Astronomische Nachrichten}\ }\textbf
  {\bibinfo {volume} {307}},\ \bibinfo {pages} {231} (\bibinfo {year}
  {1986})},\ \Eprint
  {https://arxiv.org/abs/https://onlinelibrary.wiley.com/doi/pdf/10.1002/asna.2113070403}
  {https://onlinelibrary.wiley.com/doi/pdf/10.1002/asna.2113070403}
  \BibitemShut {NoStop}%
\bibitem [{\citenamefont {Ng}\ and\ \citenamefont {Sze}(1991)}]{Ng:1991}%
  \BibitemOpen
  \bibfield  {author} {\bibinfo {author} {\bibfnamefont {K.~W.}\ \bibnamefont
  {Ng}}\ and\ \bibinfo {author} {\bibfnamefont {W.~K.}\ \bibnamefont {Sze}},\
  }\bibfield  {title} {\bibinfo {title} {Quark-hadron phase transition of the
  early universe in the nontopological soliton model},\ }\href
  {https://doi.org/10.1103/PhysRevD.43.3813} {\bibfield  {journal} {\bibinfo
  {journal} {Phys. Rev. D}\ }\textbf {\bibinfo {volume} {43}},\ \bibinfo
  {pages} {3813} (\bibinfo {year} {1991})}\BibitemShut {NoStop}%
\bibitem [{\citenamefont {Borghini}\ \emph {et~al.}(2000)\citenamefont
  {Borghini}, \citenamefont {Cottingham},\ and\ \citenamefont
  {Mau}}]{Borghini_2000}%
  \BibitemOpen
  \bibfield  {author} {\bibinfo {author} {\bibfnamefont {N.}~\bibnamefont
  {Borghini}}, \bibinfo {author} {\bibfnamefont {W.~N.}\ \bibnamefont
  {Cottingham}},\ and\ \bibinfo {author} {\bibfnamefont {R.~V.}\ \bibnamefont
  {Mau}},\ }\bibfield  {title} {\bibinfo {title} {Possible cosmological
  implications of the quark-hadron phase transition},\ }\href
  {https://doi.org/10.1088/0954-3899/26/6/302} {\bibfield  {journal} {\bibinfo
  {journal} {Journal of Physics G: Nuclear and Particle Physics}\ }\textbf
  {\bibinfo {volume} {26}},\ \bibinfo {pages} {771} (\bibinfo {year}
  {2000})}\BibitemShut {NoStop}%
\bibitem [{\citenamefont {Boeckel}\ \emph {et~al.}(2011)\citenamefont
  {Boeckel}, \citenamefont {Schettler},\ and\ \citenamefont
  {Schaffner-Bielich}}]{BOECKEL2011266}%
  \BibitemOpen
  \bibfield  {author} {\bibinfo {author} {\bibfnamefont {T.}~\bibnamefont
  {Boeckel}}, \bibinfo {author} {\bibfnamefont {S.}~\bibnamefont {Schettler}},\
  and\ \bibinfo {author} {\bibfnamefont {J.}~\bibnamefont
  {Schaffner-Bielich}},\ }\bibfield  {title} {\bibinfo {title} {The
  cosmological qcd phase transition revisited},\ }\href
  {https://doi.org/https://doi.org/10.1016/j.ppnp.2011.01.017} {\bibfield
  {journal} {\bibinfo  {journal} {Progress in Particle and Nuclear Physics}\
  }\textbf {\bibinfo {volume} {66}},\ \bibinfo {pages} {266} (\bibinfo {year}
  {2011})},\ \bibinfo {note} {particle and Nuclear Astrophysics}\BibitemShut
  {NoStop}%
\bibitem [{\citenamefont {Buballa}\ and\ \citenamefont
  {Carignano}(2015)}]{Buballa:2014tba}%
  \BibitemOpen
  \bibfield  {author} {\bibinfo {author} {\bibfnamefont {M.}~\bibnamefont
  {Buballa}}\ and\ \bibinfo {author} {\bibfnamefont {S.}~\bibnamefont
  {Carignano}},\ }\bibfield  {title} {\bibinfo {title} {{Inhomogeneous chiral
  condensates}},\ }\href {https://doi.org/10.1016/j.ppnp.2014.11.001}
  {\bibfield  {journal} {\bibinfo  {journal} {Prog. Part. Nucl. Phys.}\
  }\textbf {\bibinfo {volume} {81}},\ \bibinfo {pages} {39} (\bibinfo {year}
  {2015})},\ \Eprint {https://arxiv.org/abs/1406.1367} {arXiv:1406.1367
  [hep-ph]} \BibitemShut {NoStop}%
\bibitem [{\citenamefont {Deryagin}\ \emph {et~al.}(1992)\citenamefont
  {Deryagin}, \citenamefont {Grigoriev},\ and\ \citenamefont
  {Rubakov}}]{Deryagin:1992rw}%
  \BibitemOpen
  \bibfield  {author} {\bibinfo {author} {\bibfnamefont {D.~V.}\ \bibnamefont
  {Deryagin}}, \bibinfo {author} {\bibfnamefont {D.~Y.}\ \bibnamefont
  {Grigoriev}},\ and\ \bibinfo {author} {\bibfnamefont {V.~A.}\ \bibnamefont
  {Rubakov}},\ }\bibfield  {title} {\bibinfo {title} {{Standing wave ground
  state in high density, zero temperature QCD at large N(c)}},\ }\href
  {https://doi.org/10.1142/S0217751X92000302} {\bibfield  {journal} {\bibinfo
  {journal} {Int. J. Mod. Phys. A}\ }\textbf {\bibinfo {volume} {7}},\ \bibinfo
  {pages} {659} (\bibinfo {year} {1992})}\BibitemShut {NoStop}%
\bibitem [{\citenamefont {Shuster}\ and\ \citenamefont
  {Son}(2000)}]{Shuster:1999tn}%
  \BibitemOpen
  \bibfield  {author} {\bibinfo {author} {\bibfnamefont {E.}~\bibnamefont
  {Shuster}}\ and\ \bibinfo {author} {\bibfnamefont {D.~T.}\ \bibnamefont
  {Son}},\ }\bibfield  {title} {\bibinfo {title} {{On finite density QCD at
  large N(c)}},\ }\href {https://doi.org/10.1016/S0550-3213(99)00615-X}
  {\bibfield  {journal} {\bibinfo  {journal} {Nucl. Phys. B}\ }\textbf
  {\bibinfo {volume} {573}},\ \bibinfo {pages} {434} (\bibinfo {year}
  {2000})},\ \Eprint {https://arxiv.org/abs/hep-ph/9905448}
  {arXiv:hep-ph/9905448} \BibitemShut {NoStop}%
\bibitem [{\citenamefont {Park}\ \emph {et~al.}(2000)\citenamefont {Park},
  \citenamefont {Rho}, \citenamefont {Wirzba},\ and\ \citenamefont
  {Zahed}}]{Park:1999bz}%
  \BibitemOpen
  \bibfield  {author} {\bibinfo {author} {\bibfnamefont {B.-Y.}\ \bibnamefont
  {Park}}, \bibinfo {author} {\bibfnamefont {M.}~\bibnamefont {Rho}}, \bibinfo
  {author} {\bibfnamefont {A.}~\bibnamefont {Wirzba}},\ and\ \bibinfo {author}
  {\bibfnamefont {I.}~\bibnamefont {Zahed}},\ }\bibfield  {title} {\bibinfo
  {title} {{Dense QCD: Overhauser or BCS pairing?}},\ }\href
  {https://doi.org/10.1103/PhysRevD.62.034015} {\bibfield  {journal} {\bibinfo
  {journal} {Phys. Rev. D}\ }\textbf {\bibinfo {volume} {62}},\ \bibinfo
  {pages} {034015} (\bibinfo {year} {2000})},\ \Eprint
  {https://arxiv.org/abs/hep-ph/9910347} {arXiv:hep-ph/9910347} \BibitemShut
  {NoStop}%
\bibitem [{\citenamefont {Nakano}\ and\ \citenamefont
  {Tatsumi}(2005)}]{Nakano:2004cd}%
  \BibitemOpen
  \bibfield  {author} {\bibinfo {author} {\bibfnamefont {E.}~\bibnamefont
  {Nakano}}\ and\ \bibinfo {author} {\bibfnamefont {T.}~\bibnamefont
  {Tatsumi}},\ }\bibfield  {title} {\bibinfo {title} {{Chiral symmetry and
  density wave in quark matter}},\ }\href
  {https://doi.org/10.1103/PhysRevD.71.114006} {\bibfield  {journal} {\bibinfo
  {journal} {Phys. Rev. D}\ }\textbf {\bibinfo {volume} {71}},\ \bibinfo
  {pages} {114006} (\bibinfo {year} {2005})},\ \Eprint
  {https://arxiv.org/abs/hep-ph/0411350} {arXiv:hep-ph/0411350} \BibitemShut
  {NoStop}%
\bibitem [{\citenamefont {Nickel}(2009)}]{Nickel:2009wj}%
  \BibitemOpen
  \bibfield  {author} {\bibinfo {author} {\bibfnamefont {D.}~\bibnamefont
  {Nickel}},\ }\bibfield  {title} {\bibinfo {title} {{Inhomogeneous phases in
  the Nambu-Jona-Lasino and quark-meson model}},\ }\href
  {https://doi.org/10.1103/PhysRevD.80.074025} {\bibfield  {journal} {\bibinfo
  {journal} {Phys. Rev. D}\ }\textbf {\bibinfo {volume} {80}},\ \bibinfo
  {pages} {074025} (\bibinfo {year} {2009})},\ \Eprint
  {https://arxiv.org/abs/0906.5295} {arXiv:0906.5295 [hep-ph]} \BibitemShut
  {NoStop}%
\bibitem [{\citenamefont {Frolov}\ \emph {et~al.}(2010)\citenamefont {Frolov},
  \citenamefont {Zhukovsky},\ and\ \citenamefont {Klimenko}}]{Frolov:2010wn}%
  \BibitemOpen
  \bibfield  {author} {\bibinfo {author} {\bibfnamefont {I.~E.}\ \bibnamefont
  {Frolov}}, \bibinfo {author} {\bibfnamefont {V.~C.}\ \bibnamefont
  {Zhukovsky}},\ and\ \bibinfo {author} {\bibfnamefont {K.~G.}\ \bibnamefont
  {Klimenko}},\ }\bibfield  {title} {\bibinfo {title} {{Chiral density waves in
  quark matter within the Nambu-Jona-Lasinio model in an external magnetic
  field}},\ }\href {https://doi.org/10.1103/PhysRevD.82.076002} {\bibfield
  {journal} {\bibinfo  {journal} {Phys. Rev. D}\ }\textbf {\bibinfo {volume}
  {82}},\ \bibinfo {pages} {076002} (\bibinfo {year} {2010})},\ \Eprint
  {https://arxiv.org/abs/1007.2984} {arXiv:1007.2984 [hep-ph]} \BibitemShut
  {NoStop}%
\bibitem [{\citenamefont {Heinz}\ \emph {et~al.}(2015)\citenamefont {Heinz},
  \citenamefont {Giacosa},\ and\ \citenamefont {Rischke}}]{Heinz:2013hza}%
  \BibitemOpen
  \bibfield  {author} {\bibinfo {author} {\bibfnamefont {A.}~\bibnamefont
  {Heinz}}, \bibinfo {author} {\bibfnamefont {F.}~\bibnamefont {Giacosa}},\
  and\ \bibinfo {author} {\bibfnamefont {D.~H.}\ \bibnamefont {Rischke}},\
  }\bibfield  {title} {\bibinfo {title} {{Chiral density wave in nuclear
  matter}},\ }\href {https://doi.org/10.1016/j.nuclphysa.2014.09.027}
  {\bibfield  {journal} {\bibinfo  {journal} {Nucl. Phys. A}\ }\textbf
  {\bibinfo {volume} {933}},\ \bibinfo {pages} {34} (\bibinfo {year} {2015})},\
  \Eprint {https://arxiv.org/abs/1312.3244} {arXiv:1312.3244 [nucl-th]}
  \BibitemShut {NoStop}%
\bibitem [{\citenamefont {Carignano}\ \emph {et~al.}(2014)\citenamefont
  {Carignano}, \citenamefont {Buballa},\ and\ \citenamefont
  {Schaefer}}]{Carignano:2014jla}%
  \BibitemOpen
  \bibfield  {author} {\bibinfo {author} {\bibfnamefont {S.}~\bibnamefont
  {Carignano}}, \bibinfo {author} {\bibfnamefont {M.}~\bibnamefont {Buballa}},\
  and\ \bibinfo {author} {\bibfnamefont {B.-J.}\ \bibnamefont {Schaefer}},\
  }\bibfield  {title} {\bibinfo {title} {{Inhomogeneous phases in the
  quark-meson model with vacuum fluctuations}},\ }\href
  {https://doi.org/10.1103/PhysRevD.90.014033} {\bibfield  {journal} {\bibinfo
  {journal} {Phys. Rev. D}\ }\textbf {\bibinfo {volume} {90}},\ \bibinfo
  {pages} {014033} (\bibinfo {year} {2014})},\ \Eprint
  {https://arxiv.org/abs/1404.0057} {arXiv:1404.0057 [hep-ph]} \BibitemShut
  {NoStop}%
\bibitem [{\citenamefont {Adhikari}\ \emph {et~al.}(2017)\citenamefont
  {Adhikari}, \citenamefont {Andersen},\ and\ \citenamefont
  {Kneschke}}]{Adhikari:2017ydi}%
  \BibitemOpen
  \bibfield  {author} {\bibinfo {author} {\bibfnamefont {P.}~\bibnamefont
  {Adhikari}}, \bibinfo {author} {\bibfnamefont {J.~O.}\ \bibnamefont
  {Andersen}},\ and\ \bibinfo {author} {\bibfnamefont {P.}~\bibnamefont
  {Kneschke}},\ }\bibfield  {title} {\bibinfo {title} {{Inhomogeneous chiral
  condensate in the quark-meson model}},\ }\href
  {https://doi.org/10.1103/PhysRevD.96.016013} {\bibfield  {journal} {\bibinfo
  {journal} {Phys. Rev. D}\ }\textbf {\bibinfo {volume} {96}},\ \bibinfo
  {pages} {016013} (\bibinfo {year} {2017})},\ \bibinfo {note} {[Erratum:
  Phys.Rev.D 98, 099902 (2018)]},\ \Eprint {https://arxiv.org/abs/1702.01324}
  {arXiv:1702.01324 [hep-ph]} \BibitemShut {NoStop}%
\bibitem [{\citenamefont {Buballa}\ \emph {et~al.}(2020)\citenamefont
  {Buballa}, \citenamefont {Carignano},\ and\ \citenamefont
  {Kurth}}]{Buballa:2020xaa}%
  \BibitemOpen
  \bibfield  {author} {\bibinfo {author} {\bibfnamefont {M.}~\bibnamefont
  {Buballa}}, \bibinfo {author} {\bibfnamefont {S.}~\bibnamefont {Carignano}},\
  and\ \bibinfo {author} {\bibfnamefont {L.}~\bibnamefont {Kurth}},\ }\bibfield
   {title} {\bibinfo {title} {{Inhomogeneous phases in the quark-meson model
  with explicit chiral-symmetry breaking}},\ }\href
  {https://doi.org/10.1140/epjst/e2020-000101-x} {\bibfield  {journal}
  {\bibinfo  {journal} {Eur. Phys. J. ST}\ }\textbf {\bibinfo {volume} {229}},\
  \bibinfo {pages} {3371} (\bibinfo {year} {2020})},\ \Eprint
  {https://arxiv.org/abs/2006.02133} {arXiv:2006.02133 [hep-ph]} \BibitemShut
  {NoStop}%
\bibitem [{\citenamefont {Ferrer}\ and\ \citenamefont {de~la
  Incera}(2021)}]{Ferrer:2021mpq}%
  \BibitemOpen
  \bibfield  {author} {\bibinfo {author} {\bibfnamefont {E.~J.}\ \bibnamefont
  {Ferrer}}\ and\ \bibinfo {author} {\bibfnamefont {V.}~\bibnamefont {de~la
  Incera}},\ }\bibfield  {title} {\bibinfo {title} {{Magnetic Dual Chiral
  Density Wave: A Candidate Quark Matter Phase for the Interior of Neutron
  Stars}},\ }\href {https://doi.org/10.3390/universe7120458} {\bibfield
  {journal} {\bibinfo  {journal} {Universe}\ }\textbf {\bibinfo {volume} {7}},\
  \bibinfo {pages} {458} (\bibinfo {year} {2021})},\ \Eprint
  {https://arxiv.org/abs/2201.04032} {arXiv:2201.04032 [hep-ph]} \BibitemShut
  {NoStop}%
\bibitem [{\citenamefont {Tabatabaee~Mehr}(2023)}]{TabatabaeeMehr:2023tpt}%
  \BibitemOpen
  \bibfield  {author} {\bibinfo {author} {\bibfnamefont {S.~M.~A.}\
  \bibnamefont {Tabatabaee~Mehr}},\ }\bibfield  {title} {\bibinfo {title}
  {{Chiral symmetry breaking and phase diagram of dual chiral density wave in a
  rotating quark matter}},\ }\href
  {https://doi.org/10.1103/PhysRevD.108.094042} {\bibfield  {journal} {\bibinfo
   {journal} {Phys. Rev. D}\ }\textbf {\bibinfo {volume} {108}},\ \bibinfo
  {pages} {094042} (\bibinfo {year} {2023})},\ \Eprint
  {https://arxiv.org/abs/2306.11753} {arXiv:2306.11753 [nucl-th]} \BibitemShut
  {NoStop}%
\bibitem [{\citenamefont {Pitsinigkos}\ and\ \citenamefont
  {Schmitt}(2024)}]{Pitsinigkos:2023xee}%
  \BibitemOpen
  \bibfield  {author} {\bibinfo {author} {\bibfnamefont {S.}~\bibnamefont
  {Pitsinigkos}}\ and\ \bibinfo {author} {\bibfnamefont {A.}~\bibnamefont
  {Schmitt}},\ }\bibfield  {title} {\bibinfo {title} {{Chiral crossover versus
  chiral density wave in dense nuclear matter}},\ }\href
  {https://doi.org/10.1103/PhysRevD.109.014024} {\bibfield  {journal} {\bibinfo
   {journal} {Phys. Rev. D}\ }\textbf {\bibinfo {volume} {109}},\ \bibinfo
  {pages} {014024} (\bibinfo {year} {2024})},\ \Eprint
  {https://arxiv.org/abs/2309.01603} {arXiv:2309.01603 [nucl-th]} \BibitemShut
  {NoStop}%
\bibitem [{\citenamefont {Papadopoulos}\ and\ \citenamefont
  {Schmitt}(2025{\natexlab{a}})}]{Papadopoulos:2024agt}%
  \BibitemOpen
  \bibfield  {author} {\bibinfo {author} {\bibfnamefont {O.}~\bibnamefont
  {Papadopoulos}}\ and\ \bibinfo {author} {\bibfnamefont {A.}~\bibnamefont
  {Schmitt}},\ }\bibfield  {title} {\bibinfo {title} {{How neutron star
  properties disfavor a nuclear chiral density wave}},\ }\href
  {https://doi.org/10.1103/PhysRevD.111.034010} {\bibfield  {journal} {\bibinfo
   {journal} {Phys. Rev. D}\ }\textbf {\bibinfo {volume} {111}},\ \bibinfo
  {pages} {034010} (\bibinfo {year} {2025}{\natexlab{a}})},\ \Eprint
  {https://arxiv.org/abs/2411.08023} {arXiv:2411.08023 [nucl-th]} \BibitemShut
  {NoStop}%
\bibitem [{\citenamefont {Papadopoulos}\ and\ \citenamefont
  {Schmitt}(2025{\natexlab{b}})}]{Papadopoulos:2025uig}%
  \BibitemOpen
  \bibfield  {author} {\bibinfo {author} {\bibfnamefont {O.}~\bibnamefont
  {Papadopoulos}}\ and\ \bibinfo {author} {\bibfnamefont {A.}~\bibnamefont
  {Schmitt}},\ }\bibfield  {title} {\bibinfo {title} {{Nuclear chiral density
  wave in neutron stars?}},\ }\href
  {https://doi.org/10.1016/j.jspc.2025.100221} {\bibfield  {journal} {\bibinfo
  {journal} {J. Subatomic Part. Cosmol.}\ }\textbf {\bibinfo {volume} {4}},\
  \bibinfo {pages} {100221} (\bibinfo {year} {2025}{\natexlab{b}})},\ \Eprint
  {https://arxiv.org/abs/2509.10135} {arXiv:2509.10135 [nucl-th]} \BibitemShut
  {NoStop}%
\bibitem [{\citenamefont {Schon}\ and\ \citenamefont
  {Thies}(2000)}]{Schon:2000he}%
  \BibitemOpen
  \bibfield  {author} {\bibinfo {author} {\bibfnamefont {V.}~\bibnamefont
  {Schon}}\ and\ \bibinfo {author} {\bibfnamefont {M.}~\bibnamefont {Thies}},\
  }\bibfield  {title} {\bibinfo {title} {{Emergence of Skyrme crystal in
  Gross-Neveu and 't Hooft models at finite density}},\ }\href
  {https://doi.org/10.1103/PhysRevD.62.096002} {\bibfield  {journal} {\bibinfo
  {journal} {Phys. Rev. D}\ }\textbf {\bibinfo {volume} {62}},\ \bibinfo
  {pages} {096002} (\bibinfo {year} {2000})},\ \Eprint
  {https://arxiv.org/abs/hep-th/0003195} {arXiv:hep-th/0003195} \BibitemShut
  {NoStop}%
\bibitem [{\citenamefont {McLerran}\ and\ \citenamefont
  {Pisarski}(2007)}]{McLerran:2007qj}%
  \BibitemOpen
  \bibfield  {author} {\bibinfo {author} {\bibfnamefont {L.}~\bibnamefont
  {McLerran}}\ and\ \bibinfo {author} {\bibfnamefont {R.~D.}\ \bibnamefont
  {Pisarski}},\ }\bibfield  {title} {\bibinfo {title} {{Phases of cold, dense
  quarks at large N(c)}},\ }\href
  {https://doi.org/10.1016/j.nuclphysa.2007.08.013} {\bibfield  {journal}
  {\bibinfo  {journal} {Nucl. Phys. A}\ }\textbf {\bibinfo {volume} {796}},\
  \bibinfo {pages} {83} (\bibinfo {year} {2007})},\ \Eprint
  {https://arxiv.org/abs/0706.2191} {arXiv:0706.2191 [hep-ph]} \BibitemShut
  {NoStop}%
\bibitem [{\citenamefont {Kojo}\ \emph {et~al.}(2010)\citenamefont {Kojo},
  \citenamefont {Hidaka}, \citenamefont {McLerran},\ and\ \citenamefont
  {Pisarski}}]{Kojo:2009ha}%
  \BibitemOpen
  \bibfield  {author} {\bibinfo {author} {\bibfnamefont {T.}~\bibnamefont
  {Kojo}}, \bibinfo {author} {\bibfnamefont {Y.}~\bibnamefont {Hidaka}},
  \bibinfo {author} {\bibfnamefont {L.}~\bibnamefont {McLerran}},\ and\
  \bibinfo {author} {\bibfnamefont {R.~D.}\ \bibnamefont {Pisarski}},\
  }\bibfield  {title} {\bibinfo {title} {{Quarkyonic Chiral Spirals}},\ }\href
  {https://doi.org/10.1016/j.nuclphysa.2010.05.053} {\bibfield  {journal}
  {\bibinfo  {journal} {Nucl. Phys. A}\ }\textbf {\bibinfo {volume} {843}},\
  \bibinfo {pages} {37} (\bibinfo {year} {2010})},\ \Eprint
  {https://arxiv.org/abs/0912.3800} {arXiv:0912.3800 [hep-ph]} \BibitemShut
  {NoStop}%
\bibitem [{\citenamefont {Tatsumi}\ and\ \citenamefont
  {Muto}(2014)}]{Tatsumi:2014cea}%
  \BibitemOpen
  \bibfield  {author} {\bibinfo {author} {\bibfnamefont {T.}~\bibnamefont
  {Tatsumi}}\ and\ \bibinfo {author} {\bibfnamefont {T.}~\bibnamefont {Muto}},\
  }\bibfield  {title} {\bibinfo {title} {{Quark beta decay in the inhomogeneous
  chiral phase and cooling of compact stars}},\ }\href
  {https://doi.org/10.1103/PhysRevD.89.103005} {\bibfield  {journal} {\bibinfo
  {journal} {Phys. Rev. D}\ }\textbf {\bibinfo {volume} {89}},\ \bibinfo
  {pages} {103005} (\bibinfo {year} {2014})},\ \Eprint
  {https://arxiv.org/abs/1403.1927} {arXiv:1403.1927 [nucl-th]} \BibitemShut
  {NoStop}%
\bibitem [{\citenamefont {Buballa}\ and\ \citenamefont
  {Carignano}(2016)}]{Buballa:2015awa}%
  \BibitemOpen
  \bibfield  {author} {\bibinfo {author} {\bibfnamefont {M.}~\bibnamefont
  {Buballa}}\ and\ \bibinfo {author} {\bibfnamefont {S.}~\bibnamefont
  {Carignano}},\ }\bibfield  {title} {\bibinfo {title} {{Inhomogeneous chiral
  symmetry breaking in dense neutron-star matter}},\ }\href
  {https://doi.org/10.1140/epja/i2016-16057-6} {\bibfield  {journal} {\bibinfo
  {journal} {Eur. Phys. J. A}\ }\textbf {\bibinfo {volume} {52}},\ \bibinfo
  {pages} {57} (\bibinfo {year} {2016})},\ \Eprint
  {https://arxiv.org/abs/1508.04361} {arXiv:1508.04361 [nucl-th]} \BibitemShut
  {NoStop}%
\bibitem [{\citenamefont {Carignano}\ \emph {et~al.}(2015)\citenamefont
  {Carignano}, \citenamefont {Ferrer}, \citenamefont {de~la Incera},\ and\
  \citenamefont {Paulucci}}]{Carignano:2015kda}%
  \BibitemOpen
  \bibfield  {author} {\bibinfo {author} {\bibfnamefont {S.}~\bibnamefont
  {Carignano}}, \bibinfo {author} {\bibfnamefont {E.~J.}\ \bibnamefont
  {Ferrer}}, \bibinfo {author} {\bibfnamefont {V.}~\bibnamefont {de~la
  Incera}},\ and\ \bibinfo {author} {\bibfnamefont {L.}~\bibnamefont
  {Paulucci}},\ }\bibfield  {title} {\bibinfo {title} {{Crystalline chiral
  condensates as a component of compact stars}},\ }\href
  {https://doi.org/10.1103/PhysRevD.92.105018} {\bibfield  {journal} {\bibinfo
  {journal} {Phys. Rev. D}\ }\textbf {\bibinfo {volume} {92}},\ \bibinfo
  {pages} {105018} (\bibinfo {year} {2015})},\ \Eprint
  {https://arxiv.org/abs/1505.05094} {arXiv:1505.05094 [nucl-th]} \BibitemShut
  {NoStop}%
\bibitem [{\citenamefont {Guenther}(2021)}]{Guenther:2020jwe}%
  \BibitemOpen
  \bibfield  {author} {\bibinfo {author} {\bibfnamefont {J.~N.}\ \bibnamefont
  {Guenther}},\ }\bibfield  {title} {\bibinfo {title} {{Overview of the QCD
  phase diagram: Recent progress from the lattice}},\ }\href
  {https://doi.org/10.1140/epja/s10050-021-00354-6} {\bibfield  {journal}
  {\bibinfo  {journal} {Eur. Phys. J. A}\ }\textbf {\bibinfo {volume} {57}},\
  \bibinfo {pages} {136} (\bibinfo {year} {2021})},\ \Eprint
  {https://arxiv.org/abs/2010.15503} {arXiv:2010.15503 [hep-lat]} \BibitemShut
  {NoStop}%
\bibitem [{\citenamefont {Boguta}(1983)}]{Boguta:1982wr}%
  \BibitemOpen
  \bibfield  {author} {\bibinfo {author} {\bibfnamefont {J.}~\bibnamefont
  {Boguta}},\ }\bibfield  {title} {\bibinfo {title} {{A SATURATING CHIRAL FIELD
  THEORY OF NUCLEAR MATTER}},\ }\href
  {https://doi.org/10.1016/0370-2693(83)90617-2} {\bibfield  {journal}
  {\bibinfo  {journal} {Phys. Lett. B}\ }\textbf {\bibinfo {volume} {120}},\
  \bibinfo {pages} {34} (\bibinfo {year} {1983})}\BibitemShut {NoStop}%
\bibitem [{\citenamefont {Floerchinger}\ and\ \citenamefont
  {Wetterich}(2012)}]{Floerchinger:2012xd}%
  \BibitemOpen
  \bibfield  {author} {\bibinfo {author} {\bibfnamefont {S.}~\bibnamefont
  {Floerchinger}}\ and\ \bibinfo {author} {\bibfnamefont {C.}~\bibnamefont
  {Wetterich}},\ }\bibfield  {title} {\bibinfo {title} {{Chemical freeze-out in
  heavy ion collisions at large baryon densities}},\ }\href
  {https://doi.org/10.1016/j.nuclphysa.2012.07.009} {\bibfield  {journal}
  {\bibinfo  {journal} {Nucl. Phys. A}\ }\textbf {\bibinfo {volume}
  {890-891}},\ \bibinfo {pages} {11} (\bibinfo {year} {2012})},\ \Eprint
  {https://arxiv.org/abs/1202.1671} {arXiv:1202.1671 [nucl-th]} \BibitemShut
  {NoStop}%
\bibitem [{\citenamefont {Drews}\ \emph {et~al.}(2013)\citenamefont {Drews},
  \citenamefont {Hell}, \citenamefont {Klein},\ and\ \citenamefont
  {Weise}}]{Drews:2013hha}%
  \BibitemOpen
  \bibfield  {author} {\bibinfo {author} {\bibfnamefont {M.}~\bibnamefont
  {Drews}}, \bibinfo {author} {\bibfnamefont {T.}~\bibnamefont {Hell}},
  \bibinfo {author} {\bibfnamefont {B.}~\bibnamefont {Klein}},\ and\ \bibinfo
  {author} {\bibfnamefont {W.}~\bibnamefont {Weise}},\ }\bibfield  {title}
  {\bibinfo {title} {{Thermodynamic phases and mesonic fluctuations in a chiral
  nucleon-meson model}},\ }\href {https://doi.org/10.1103/PhysRevD.88.096011}
  {\bibfield  {journal} {\bibinfo  {journal} {Phys. Rev. D}\ }\textbf {\bibinfo
  {volume} {88}},\ \bibinfo {pages} {096011} (\bibinfo {year} {2013})},\
  \Eprint {https://arxiv.org/abs/1308.5596} {arXiv:1308.5596 [hep-ph]}
  \BibitemShut {NoStop}%
\bibitem [{\citenamefont {Drews}\ and\ \citenamefont
  {Weise}(2015)}]{Drews:2014spa}%
  \BibitemOpen
  \bibfield  {author} {\bibinfo {author} {\bibfnamefont {M.}~\bibnamefont
  {Drews}}\ and\ \bibinfo {author} {\bibfnamefont {W.}~\bibnamefont {Weise}},\
  }\bibfield  {title} {\bibinfo {title} {{From asymmetric nuclear matter to
  neutron stars: a functional renormalization group study}},\ }\href
  {https://doi.org/10.1103/PhysRevC.91.035802} {\bibfield  {journal} {\bibinfo
  {journal} {Phys. Rev. C}\ }\textbf {\bibinfo {volume} {91}},\ \bibinfo
  {pages} {035802} (\bibinfo {year} {2015})},\ \Eprint
  {https://arxiv.org/abs/1412.7655} {arXiv:1412.7655 [nucl-th]} \BibitemShut
  {NoStop}%
\bibitem [{\citenamefont {Campbell}\ \emph {et~al.}(1990)\citenamefont
  {Campbell}, \citenamefont {Ellis},\ and\ \citenamefont
  {Olive}}]{Campbell:1989gh}%
  \BibitemOpen
  \bibfield  {author} {\bibinfo {author} {\bibfnamefont {B.~A.}\ \bibnamefont
  {Campbell}}, \bibinfo {author} {\bibfnamefont {J.~R.}\ \bibnamefont
  {Ellis}},\ and\ \bibinfo {author} {\bibfnamefont {K.~A.}\ \bibnamefont
  {Olive}},\ }\bibfield  {title} {\bibinfo {title} {{EFFECTIVE LAGRANGIAN
  APPROACH TO QCD PHASE TRANSITIONS}},\ }\href
  {https://doi.org/10.1016/0370-2693(90)91973-F} {\bibfield  {journal}
  {\bibinfo  {journal} {Phys. Lett. B}\ }\textbf {\bibinfo {volume} {235}},\
  \bibinfo {pages} {325} (\bibinfo {year} {1990})}\BibitemShut {NoStop}%
\bibitem [{\citenamefont {Giordano}\ \emph {et~al.}(2021)\citenamefont
  {Giordano}, \citenamefont {Kapas}, \citenamefont {Katz}, \citenamefont
  {Nogradi},\ and\ \citenamefont {Pasztor}}]{Giordano:2020huj}%
  \BibitemOpen
  \bibfield  {author} {\bibinfo {author} {\bibfnamefont {M.}~\bibnamefont
  {Giordano}}, \bibinfo {author} {\bibfnamefont {K.}~\bibnamefont {Kapas}},
  \bibinfo {author} {\bibfnamefont {S.~D.}\ \bibnamefont {Katz}}, \bibinfo
  {author} {\bibfnamefont {D.}~\bibnamefont {Nogradi}},\ and\ \bibinfo {author}
  {\bibfnamefont {A.}~\bibnamefont {Pasztor}},\ }\bibfield  {title} {\bibinfo
  {title} {{Towards a reliable lower bound on the location of the critical
  endpoint}},\ }\href {https://doi.org/10.1016/j.nuclphysa.2020.121986}
  {\bibfield  {journal} {\bibinfo  {journal} {Nucl. Phys. A}\ }\textbf
  {\bibinfo {volume} {1005}},\ \bibinfo {pages} {121986} (\bibinfo {year}
  {2021})},\ \Eprint {https://arxiv.org/abs/2004.07066} {arXiv:2004.07066
  [hep-lat]} \BibitemShut {NoStop}%
\bibitem [{\citenamefont {Borsanyi}\ \emph {et~al.}(2025)\citenamefont
  {Borsanyi}, \citenamefont {Fodor}, \citenamefont {Guenther}, \citenamefont
  {Parotto}, \citenamefont {Pasztor}, \citenamefont {Ratti}, \citenamefont
  {Vovchenko},\ and\ \citenamefont {Wong}}]{Borsanyi:2025dyp}%
  \BibitemOpen
  \bibfield  {author} {\bibinfo {author} {\bibfnamefont {S.}~\bibnamefont
  {Borsanyi}}, \bibinfo {author} {\bibfnamefont {Z.}~\bibnamefont {Fodor}},
  \bibinfo {author} {\bibfnamefont {J.~N.}\ \bibnamefont {Guenther}}, \bibinfo
  {author} {\bibfnamefont {P.}~\bibnamefont {Parotto}}, \bibinfo {author}
  {\bibfnamefont {A.}~\bibnamefont {Pasztor}}, \bibinfo {author} {\bibfnamefont
  {C.}~\bibnamefont {Ratti}}, \bibinfo {author} {\bibfnamefont
  {V.}~\bibnamefont {Vovchenko}},\ and\ \bibinfo {author} {\bibfnamefont
  {C.~H.}\ \bibnamefont {Wong}},\ }\bibfield  {title} {\bibinfo {title}
  {{Lattice QCD constraints on the critical point from an improved precision
  equation of state}},\ }\href {https://doi.org/10.1103/rj6r-dmg9} {\bibfield
  {journal} {\bibinfo  {journal} {Phys. Rev. D}\ }\textbf {\bibinfo {volume}
  {112}},\ \bibinfo {pages} {L111505} (\bibinfo {year} {2025})},\ \Eprint
  {https://arxiv.org/abs/2502.10267} {arXiv:2502.10267 [hep-lat]} \BibitemShut
  {NoStop}%
\bibitem [{\citenamefont {Fraga}\ \emph {et~al.}(2011)\citenamefont {Fraga},
  \citenamefont {Palhares},\ and\ \citenamefont {Sorensen}}]{Fraga:2011hi}%
  \BibitemOpen
  \bibfield  {author} {\bibinfo {author} {\bibfnamefont {E.~S.}\ \bibnamefont
  {Fraga}}, \bibinfo {author} {\bibfnamefont {L.~F.}\ \bibnamefont
  {Palhares}},\ and\ \bibinfo {author} {\bibfnamefont {P.}~\bibnamefont
  {Sorensen}},\ }\bibfield  {title} {\bibinfo {title} {{Finite-size scaling as
  a tool in the search for the QCD critical point in heavy ion data}},\ }\href
  {https://doi.org/10.1103/PhysRevC.84.011903} {\bibfield  {journal} {\bibinfo
  {journal} {Phys. Rev. C}\ }\textbf {\bibinfo {volume} {84}},\ \bibinfo
  {pages} {011903} (\bibinfo {year} {2011})},\ \Eprint
  {https://arxiv.org/abs/1104.3755} {arXiv:1104.3755 [hep-ph]} \BibitemShut
  {NoStop}%
\bibitem [{\citenamefont {Sorensen}\ and\ \citenamefont
  {Sorensen}(2024)}]{Sorensen:2024mry}%
  \BibitemOpen
  \bibfield  {author} {\bibinfo {author} {\bibfnamefont {A.}~\bibnamefont
  {Sorensen}}\ and\ \bibinfo {author} {\bibfnamefont {P.}~\bibnamefont
  {Sorensen}},\ }\bibfield  {title} {\bibinfo {title} {{Locating the critical
  point for the hadron to quark-gluon plasma phase transition from finite-size
  scaling of proton cumulants in heavy-ion collisions}},\ }\href@noop {} {\
  (\bibinfo {year} {2024})},\ \Eprint {https://arxiv.org/abs/2405.10278}
  {arXiv:2405.10278 [nucl-th]} \BibitemShut {NoStop}%
\bibitem [{\citenamefont {Lacey}(2026)}]{Lacey:2026rhc}%
  \BibitemOpen
  \bibfield  {author} {\bibinfo {author} {\bibfnamefont {R.~A.}\ \bibnamefont
  {Lacey}},\ }\bibfield  {title} {\bibinfo {title} {{Finite-Size Scaling of
  Net-Proton Cumulants in Heavy-Ion Collisions: Remarks on the Interpretation
  of a Recent Analysis}},\ }\href@noop {} {\  (\bibinfo {year} {2026})},\
  \Eprint {https://arxiv.org/abs/2603.10399} {arXiv:2603.10399 [nucl-th]}
  \BibitemShut {NoStop}%
\bibitem [{\citenamefont {Glendenning}(1982)}]{Glendenning:1982nc}%
  \BibitemOpen
  \bibfield  {author} {\bibinfo {author} {\bibfnamefont {N.~K.}\ \bibnamefont
  {Glendenning}},\ }\bibfield  {title} {\bibinfo {title} {{THE HYPERON
  COMPOSITION OF NEUTRON STARS}},\ }\href
  {https://doi.org/10.1016/0370-2693(82)90078-8} {\bibfield  {journal}
  {\bibinfo  {journal} {Phys. Lett. B}\ }\textbf {\bibinfo {volume} {114}},\
  \bibinfo {pages} {392} (\bibinfo {year} {1982})}\BibitemShut {NoStop}%
\bibitem [{\citenamefont {{Glendenning}}(1985)}]{1985ApJ...293..470G}%
  \BibitemOpen
  \bibfield  {author} {\bibinfo {author} {\bibfnamefont {N.~K.}\ \bibnamefont
  {{Glendenning}}},\ }\bibfield  {title} {\bibinfo {title} {{Neutron stars are
  giant hypernuclei ?}},\ }\href {https://doi.org/10.1086/163253} {\bibfield
  {journal} {\bibinfo  {journal} {\apj}\ }\textbf {\bibinfo {volume} {293}},\
  \bibinfo {pages} {470} (\bibinfo {year} {1985})}\BibitemShut {NoStop}%
\bibitem [{\citenamefont {Glendenning}(1997)}]{Glendenning:1997wn}%
  \BibitemOpen
  \bibfield  {author} {\bibinfo {author} {\bibfnamefont {N.~K.}\ \bibnamefont
  {Glendenning}},\ }\href@noop {} {\emph {\bibinfo {title} {{Compact stars:
  Nuclear physics, particle physics, and general relativity}}}}\ (\bibinfo
  {year} {1997})\BibitemShut {NoStop}%
\bibitem [{\citenamefont {Blaizot}(1980)}]{BLAIZOT1980171}%
  \BibitemOpen
  \bibfield  {author} {\bibinfo {author} {\bibfnamefont {J.}~\bibnamefont
  {Blaizot}},\ }\bibfield  {title} {\bibinfo {title} {Nuclear
  compressibilities},\ }\href
  {https://doi.org/https://doi.org/10.1016/0370-1573(80)90001-0} {\bibfield
  {journal} {\bibinfo  {journal} {Physics Reports}\ }\textbf {\bibinfo {volume}
  {64}},\ \bibinfo {pages} {171} (\bibinfo {year} {1980})}\BibitemShut
  {NoStop}%
\bibitem [{\citenamefont {Sharma}\ \emph {et~al.}(1988)\citenamefont {Sharma},
  \citenamefont {Borghols}, \citenamefont {Brandenburg}, \citenamefont {Crona},
  \citenamefont {van~der Woude},\ and\ \citenamefont
  {Harakeh}}]{PhysRevC.38.2562}%
  \BibitemOpen
  \bibfield  {author} {\bibinfo {author} {\bibfnamefont {M.~M.}\ \bibnamefont
  {Sharma}}, \bibinfo {author} {\bibfnamefont {W.~T.~A.}\ \bibnamefont
  {Borghols}}, \bibinfo {author} {\bibfnamefont {S.}~\bibnamefont
  {Brandenburg}}, \bibinfo {author} {\bibfnamefont {S.}~\bibnamefont {Crona}},
  \bibinfo {author} {\bibfnamefont {A.}~\bibnamefont {van~der Woude}},\ and\
  \bibinfo {author} {\bibfnamefont {M.~N.}\ \bibnamefont {Harakeh}},\
  }\bibfield  {title} {\bibinfo {title} {Giant monopole resonance in sn and sm
  nuclei and the compressibility of nuclear matter},\ }\href
  {https://doi.org/10.1103/PhysRevC.38.2562} {\bibfield  {journal} {\bibinfo
  {journal} {Phys. Rev. C}\ }\textbf {\bibinfo {volume} {38}},\ \bibinfo
  {pages} {2562} (\bibinfo {year} {1988})}\BibitemShut {NoStop}%
\bibitem [{\citenamefont {Navas}\ \emph {et~al.}(2024)\citenamefont {Navas}
  \emph {et~al.}}]{ParticleDataGroup:2024cfk}%
  \BibitemOpen
  \bibfield  {author} {\bibinfo {author} {\bibfnamefont {S.}~\bibnamefont
  {Navas}} \emph {et~al.} (\bibinfo {collaboration} {Particle Data Group}),\
  }\bibfield  {title} {\bibinfo {title} {{Review of particle physics}},\ }\href
  {https://doi.org/10.1103/PhysRevD.110.030001} {\bibfield  {journal} {\bibinfo
   {journal} {Phys. Rev. D}\ }\textbf {\bibinfo {volume} {110}},\ \bibinfo
  {pages} {030001} (\bibinfo {year} {2024})}\BibitemShut {NoStop}%
\bibitem [{\citenamefont {{National Research
  Council}}(1999)}]{1999nap..book.6288N}%
  \BibitemOpen
  \bibfield  {author} {\bibinfo {author} {\bibnamefont {{National Research
  Council}}},\ }\bibfield  {title} {\bibinfo {title} {{Nuclear Physics: The
  Core of Matter, The Fuel of Stars}},\ }in\ \href
  {https://doi.org/10.17226/6288} {\emph {\bibinfo {booktitle} {National
  Research Council. 1999. Nuclear Physics: The Core of Matter}}}\ (\bibinfo
  {year} {1999})\ p.\ \bibinfo {pages} {6288}\BibitemShut {NoStop}%
\end{thebibliography}%

\end{document}